\documentclass[11pt]{article}
\usepackage{setspace}
\usepackage{graphicx}
\usepackage[space]{grffile}
\usepackage{amsmath}
\usepackage{amsthm}
\usepackage[sort,comma,authoryear]{natbib}
\usepackage{mathtools}
\usepackage{bm}
\usepackage{amssymb}
\usepackage{threeparttable}
\DeclarePairedDelimiter{\abs}{\lvert}{\rvert}
\usepackage{booktabs}
\usepackage{array}
\usepackage[english]{babel}
\usepackage{longtable}
\usepackage{subfig}
\usepackage{url}
\usepackage{ragged2e}
\usepackage{tabularx}
\usepackage{epstopdf}
\usepackage[toc,title,page]{appendix}

\usepackage{longtable}
\usepackage{threeparttable}
\usepackage{rotating}

\theoremstyle{definition}

\theoremstyle{remark}

\usepackage{tabularx}
\usepackage{changepage}
\usepackage{verbatim}
\usepackage[a4paper, total={6in, 8in}]{geometry}
\usepackage{setspace}
\usepackage{booktabs}
\usepackage{geometry}
\usepackage{multirow}
\numberwithin{equation}{section}
\usepackage{geometry}

\newcommand{\red}[1]{\textcolor{red}{#1}}
\newcommand{\tabincell}[2]{\begin{tabular}{@{}#1@{}}#2\end{tabular}}
\usepackage{hyperref}
\usepackage{natbib}
\hypersetup{
	colorlinks,
	linkcolor={blue},
	citecolor={blue},
	urlcolor={blue}
}
\usepackage[ruled]{algorithm2e}
\begin{document}

\author{
\quad  Zhengxiao Li \footnote{School of Insurance and Economics, University of International Business and Economics, Beijing, China. Email: li\_zhengxiao@uibe.edu.cn.}
\quad Fei Wang \footnote{School of Insurance and Economics, University of International Business and Economics, Beijing, China; Taikang Life Insurance Co.,Ltd., Beijing, China. 
}
\quad Zhengtang Zhao \footnote{Corresponding author: ztzhao@xmu.edu.cn. Department of Finance, School of Economics, Xiamen University, Xiamen, China.}
}

\date{}




\title{A new class of composite GBII regression models with varying threshold for modelling heavy-tailed data}


\maketitle

\begin{abstract}
The four-parameter generalized beta distribution of the second kind (GBII) has been proposed for modelling
insurance losses with heavy-tailed features.
The aim of this paper is to present a parametric composite GBII regression modelling
by splicing two GBII distributions using mode matching method.
It is designed for simultaneous modelling of small and large claims and capturing the policyholder heterogeneity by introducing the covariates into the scale parameter.
The threshold that splits two GBII distributions is allowed to vary across individuals policyholders based on their risk features.
The proposed regression modelling also contains a wide range of insurance loss distributions as the head and the tail respectively and provides the close-formed expressions for parameter estimation and model prediction.
A simulation study is conducted to show the accuracy of the proposed estimation method and the flexibility of the regressions.
Some  illustrations of the applicability of the new class of distributions and regressions are provided with a Danish fire losses data set and a Chinese medical insurance claims data set, comparing with
the results of competing models from the literature. 
\\
\\
{\bf{Article History:}}  \today
\end{abstract}
\newpage

\section{Introduction}

For many lines of insurance business, the insurance loss data often exhibits the heterogeneity,  heavy-tailedness and different tail behavior of small and large amounts.
Various statistical methods have been proposed to generalized classical loss distributions to account for the above mentioned characteristics
of the loss data, 
which are based on, but not limited to the following three methods: (1) transformation method, 
(2) mixed or mixture method, 
(3) splicing method (also known as composite model).

The transformed method
provides procedures to fit the
log-transformed distribution
to the data with heavy-tailed features,
for instance 
the skew Normal distribution \citep{azzalini2002log}, the log skew T distribution \citep{landsman2016tail}, the generalized log Moyal distribution \citep{bhati2018generalized}.
A drawback of the transformation method is that it may magnify the error of the prediction as it changes the variance structure of the data \citep{gan2018fat}.
The four-parameter GBII family as a transformed beta family
that includes many of the classical loss distributions as a special
or limiting case has been proved to be a very useful tool in the actuarial literature 
\citep{dong2013bayesian,shi2018pair}.



Mixed or mixture model
constitutes another method which deals with
modelling insurance losses with unobserved heterogeneity and heavy-tailed features.
This approach has been discussed in several publications
in non-life actuarial literature.
For example, \cite{chan2018modelling} extend the GBII family to the contaminated
GBII family based on a finite mixture method which aims to capture the bimodality and a wide range of skewness and kurtosis of insurance loss data.
Here, we refer to many other mixture and mixed models:
a finite mixture of skew Normal distributions \citep{bernardi2012skew},
a finite mixture of Erlang distributions \citep{verbelen2015fitting},
a Gamma mixture with the generalized inverse Gaussian distribution \citep{gomez2013gamma}, and
more general finite mixture models based on the Burr, Gamma, inverse Burr, inverse Gaussian, Lognormal and Weibull distributions \citep{miljkovic2016modeling}.
However, 
as the derivatives of the log-likelihood function of mixture models are complicated in the classical likelihood approach, 
the expectation-maximization (EM) techniques need to be applied in estimation procedure, which suffers 
significant challenges on the initialization
of parameter estimates.
Recently, \cite{punzo2018compound} propose a three-parameter compound distribution (also known as mixed distribution) in order to take care of specifics
such as unimodality, hump-shaped, right-skewed, and heavy tails.
The resultant density obtained by this work also does not always have
closed-form expressions, which makes the estimation more cumbersome.

The splicing method as the third approach provides a global model fit strategy 
by combining a tail fit and a distribution modelling the loss data below the threshold.
Different spliced or so-called composite models emerge depending on the univariate distribution used for the head and the tail,
see e.g. \cite{cooray2005modeling}, \cite{scollnik2007composite}, \cite{scollnik2012modeling}, \cite{bakar2015modeling} and \cite{del2017full}. 
In \cite{reynkens2017modelling}, the modal part fit is established using a mixed Erlang distribution, which can also be adapted to censored data. This then reduces the problem of selecting a specific parametric modal part component. 
In \cite{grun2019extending},
a comprehensive analysis is provided for the Danish fire losses data set by evaluating
256 composite models derived from 16 parametric distributions that are
commonly used in actuarial science.
However, the estimation procedure in \cite{grun2019extending} requires the numeric optimization, derivative calculation, as well as root finding methods as even the parameters in the density function does not have closed-form expressions when different combinations of distribution for the head and tail part are considered.

The use of covariate information in order to predict heavy-tailed
loss data through regression models has been recently become a popular topic for insurance pricing, reserving and risk measurement as traditional generalized linear modelling such as Gamma and inverse Gaussian
regression do not have specific interest in heavy-tailed modelling.
An important
contribution of this kind in non-life insurance rate making and reserving is \cite{shi2018pair} and \cite{dong2013bayesian}, in which the GBII family 
was used as a response
distribution for regression modelling. 
Although the regression modelling in transformed distributions and mixture/mixed distributions are well developed, 
for instance, 
the Burr regression \citep{beirlant1998burr},
the generalized log-Moyal regression \citep{bhati2018generalized},
the mixed exponential regression \citep{tzougas2020algorithm},
the subfamily of GBII regression \citep{li2021generalizing} and
the Phase-type regression \citep{bladt2021phase},
the use of heavy-tailed composite models
in a regression setting has not been fully established yet.
\cite{gan2018fat} propose the use of two spliced regression models for modelling the aggregate loss data directly. The interpreting the regression coefficients of the spliced models is not straightforward.
\cite{laudage2019severity} propose a composite model with a regression structure, only allowing 
the covariates to be introduced in the body of the loss distribution. 
While \cite{fung2021mixture} presents a mixture composite regression model by introducing systematic effects of covariates in body and tail of the distribution,
the threshold is still pre-determined using export information via performing extreme value analysis.
The above mentioned models also impose an additional constraint of fixed threshold when the covariates are considered, which does not reflect the  policyholder heterogeneity among the tail part of the loss data.
The recent work about varying threshold used in the composite regression modelling is 
a deep composite regression
model proposed by \cite{fissler2021deep} whose splicing point is given in terms of a quantile of the conditional claim size
distribution rather than a constant.

The aim of this paper is to introduce a general family of composite regression models with varying threshold for approximating both the modal part and the tail of heavy-tailed loss data as well as for capturing the risk heterogeneity among policyholders.
For this purpose, we first splice two GBII distribution as a head and tail part respectively by using a mode-matching method proposed by \cite{calderin2016modeling}.
This method can incorporate unrestricted mixing weights and gives a simpler derivation of the
model over the traditional continuity-differentiability method discussed in \cite{grun2019extending}. 
The new class of distributions contains a wide range of insurance loss distributions as the head and the tail respectively and is
very flexible in modelling different
shapes of distributions. 
It also provides the close-formed expressions for parameter estimation and model prediction.
Next, 
the composite GBII family is used as a response distribution for regression modelling, in which the scale parameter is modelled as a function of several covariates in a non-linear form.
The proposed regression setting is defined so that its mean
is proportional to some exponential transformation of the linear combinations of covariates.
It also gives the explicit expressions of VaR across all individuals and related to covariates when the mean does not always exist for modelling more extreme losses.
Moreover,
the threshold that splits the two GBII distribution varies across policyholders based on observed risk features, which
allows us to capture different tail behaviors among individuals.
{Finally, 
we discuss the identification problem of model parameters and present a constrained maximum likelihood estimation method by re-parameterizing the regression model to improve the stability of parameter estimation in statistical inference.
The non-linear optimization problem is solved by using the augmented Lagrange multiplier method.} 
The estimation results
are demonstrated to perform satisfactorily when the composite GBII regression models are fitted
to a simulation study and two real-world insurance data sets.

The structure of this paper is as follows. 
In Section \ref{sec: model-specifications} we 
provide a brief summary of the GBII distribution,
introduce a new class of composite GBII distributions and study some properties, such as its moments and risk measure expressions.
Regression modelling and estimation procedure are discussed in Section \ref{sec: regression}.
The advantages of the composite GBII regressions
compared to the conventional GLMs in the presence of different tail behaviors
are illustrated by a simulation study in Section \ref{sec: simulation}.
To illustrate its practical use, in Section \ref{sec: applications}, we first fit the composite GBII models
to
the well-known Danish fire insurance data-set, comparing with fits based on models from existing literature. 
Then regression modelling is discussed 
with an illustration on a Chinese medical insurance data set.
Section \ref{section:conclusion} gives some conclusions and future possible extensions.
{To make the framework and estimation algorithm more accessible to practitioners, we provide the R programming codes for implementation at \url{https://github.com/lizhengxiao/ComGBII-Regressions}.
}

\section{Model specifications}\label{sec: model-specifications}
\subsection{The GBII distribution}
Let $Y \in \mathbb{R}^+$ be the insurance loss or claim amount random variable. The density of the GBII distribution is
given by:
\begin{equation}
\label{eq: pdf-GBII}
f_{\text{GBII}}\left(y ;p,\mu, \nu, \tau \right)=\frac{\abs{p}}{B(\nu,\tau)y}\frac{\mu^{p \tau}y^{p\nu}}{(y^p+\mu^p)^{\nu+\tau}},
\end{equation}
for $y>0$, $\mu, \nu,\tau>0$ with  $B(m,n)=\int_0^{1}t^{m-1}(1-t)^{n-1}dt$ the beta function. 
{When the parameter $p <0$, \eqref{eq: pdf-GBII} admit the inverse distributions, which is obtained by making the reciprocal transformation \citep{mcdonald1995generalization}.}
The cumulative distribution function (cdf) and quantile function (qf) of the GBII distribution are given by respectively
\begin{align}
\label{cdf:GBII}
F_{\text{GBII}}\left(y;p,\mu, \nu, \tau  \right)&=I_{\nu,\tau}\left[\frac{(y/\mu)^p}{1 + (y/\mu)^p}  \right],\quad y>0,\\
F^{-1}_{\text{GBII}}(q;p,\mu, \nu, \tau  )&=\mu\left[\frac{I_{\nu,\tau}^{-1}(q)}{1- I_{\nu,\tau}^{-1}(q)}\right]^{\frac{1}{p}},\quad q \in (0,1),
\label{qf:GBII}
\end{align}
where  
$q$ is the quantile level and 
$I^{-1}_{m,n}(\cdot)$ denotes the inverse of the beta cumulative distribution function $I_{m,n}(\cdot)$  (or regularized incomplete beta function).
The mode of the GBII occurs at
\begin{equation}
\label{mode:GBII}
y_0=\mu\left(\frac{p\nu-1}{p\tau+1}\right)^{1/p},\quad \text{if}\quad p\nu >1,
\end{equation}
and at zero otherwise.

{With four parameters, the GBII distribution is very flexible to model skewed and heavy-tailed data.
The parameter
$p$ impacts the peakedness of the density, 
whereas $\mu$ is basically a scale parameter,
and $\nu$ and $\tau$  control the shape and skewness \citep{mcdonald1995generalization} \footnote{
	$\mu$ is scale parameter and $p,\nu,\tau$ are the shape parameters.}.}
Moreover, the GBII density is regularly varying at infinity with index $-p\tau-1$ and regularly varying at the origin with index $-p\nu-1$, which implies that 
all three shape parameters control the tail behavior of the distribution.

The $h$-th moments for the GBII exist only for $-p\nu<h<p\tau$ which are given by
\[
\mathbb{E}\left[Y^h;p,\mu, \nu, \tau  \right]=\frac{\mu^h B\left(\nu+h/p,\tau-h/p\right)}{B\left(\nu,\tau\right)}.
\]

Concerning the $h$-th incomplete (conditional) moments of the GBII distribution given $y  \le s$ and $y > s$ for any positive value $s$, one finds
\begin{equation}
		\label{eq: incomplete-moments-1}
	\mathbb{E}\left[Y^h;p,\mu, \nu, \tau|Y\le s \right]=
	\mu^h\frac{B(\nu+h/p,\tau-h/p)}{B(\nu,\tau)}\frac{I_{\nu+h/p,\tau-h/p}\left[\frac{(s/\mu)^{p}}{1+(s/\mu)^{p}}\right]}{I_{\nu,\tau}\left[\frac{(s/\mu)^{p}}{1+(s/\mu)^{p}}\right]},
\end{equation}
and
 \begin{equation}
 	\label{eq: incomplete-moments-2}
 	\mathbb{E}\left[Y^h;p,\mu, \nu, \tau|Y >  s \right]=
 	\mu^h\frac{B(\nu+h/p,\tau-h/p)}{B(\nu,\tau)}\frac{1-I_{\nu+h/p,\tau-h/p}\left[\frac{(s/\mu)^{p}}{1+(s/\mu)^{p}}\right]}{1-I_{\nu,\tau}\left[\frac{(s/\mu)^{p}}{1+(s/\mu)^{p}}\right]}.
 \end{equation}

\begin{figure}[tp!]
	\centering
	\includegraphics[scale=0.6]{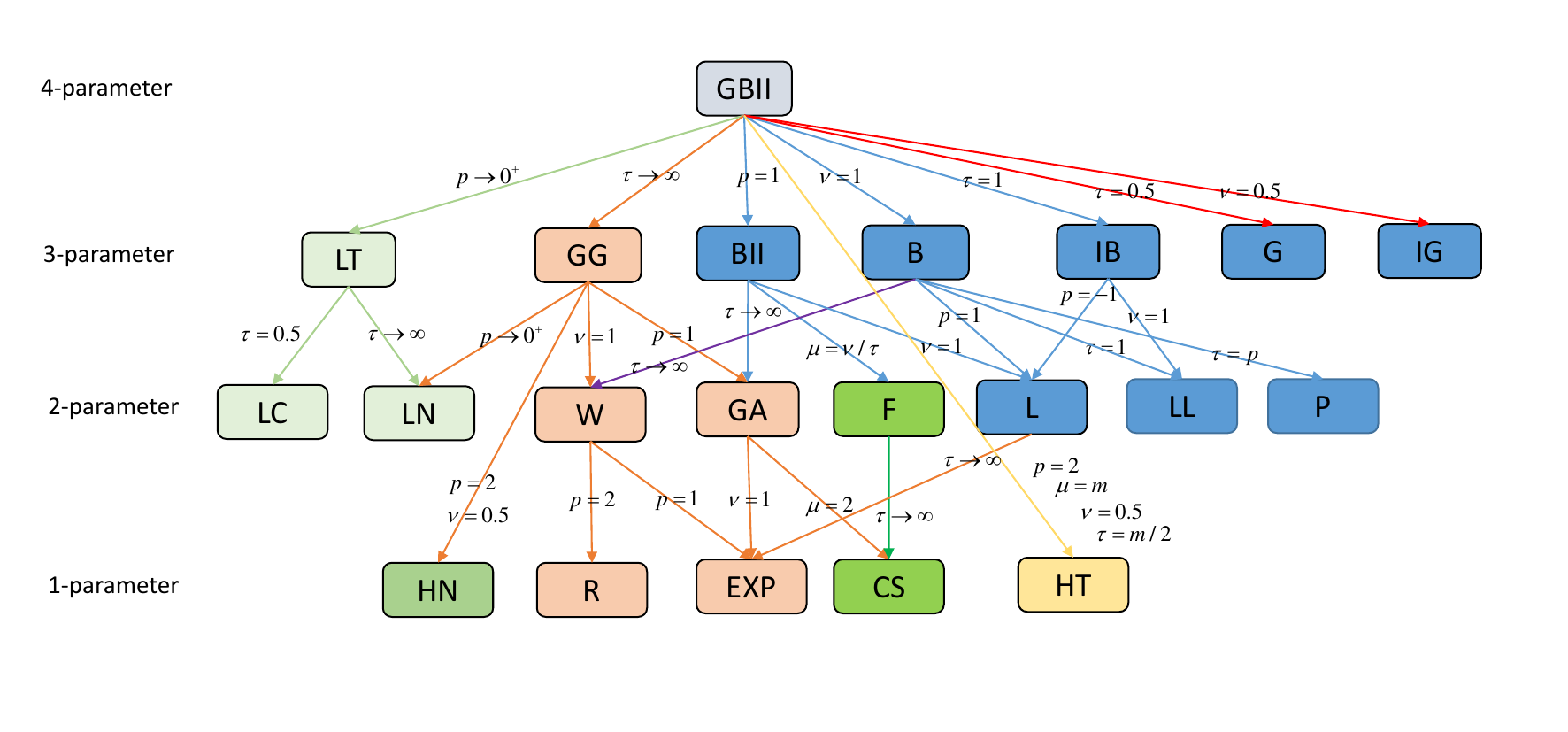}	
	\caption{Family of GBII distribution.}
	\label{fig: gb2}
\end{figure}

The relationship of GBII distribution with
many popular distributions is summarized in Figure \ref{fig: gb2}.
Figure \ref{fig: gb2} shows that the special cases of GBII distribution include three-parameter distributions of log-T (LT), generalized Gamma (GG),
beta distribution of the second kind (BII), Burr (B, also known as Singh-Maddala distribution)
and inverse Burr (IB), GLMGA (G) and inverse GLMGA (IG) distribution \footnote{
The GLMGA distribution is proposed by \cite{li2021generalizing} by mixing a generalized log-Moyal distribution (GlogM)  \citep{bhati2018generalized} with the gamma distribution. 
It is a subfamily of the GBII and belongs to the Pareto-type distribution that can be used to accommodate the extreme risks and capture both tail and modal parts of heavy-tailed insurance data. 
It occupies an interesting position in between the popular GBII model and the Lomax model.
{Figure \ref{fig: gb2} extends the Figure in \cite{chan2018modelling} by adding this new heavy-tailed distribution to the GBII family.}
},
the two-parameter distributions of log-Cauchy (LC), log-Normal (LN),
Weibull (W), Gamma (GA), Variance Ratio (F), Lomax (L),  Loglogistic (LL), Paralogistic (P)
and the one-parameter distributions of half-Normal (HN), Rayleigh (R), Exponential (EXP), Chisquare (CS) and half-T (HT) distributions.

\subsection{A new class of composite GBII models}
Given that the GBII distribution is unimodal when the parameters $p\nu>1$ and is designed to model both the light-tailed and heavy-tailed data,
a new class of 
composite models with two GBII distributions as a head and tail fit respectively
can be derived via the mode matching procedure as discussed in \cite{calderin2016modeling}. 
This method can overcome the drawbacks of the usually used splicing method  based on 
the continuity and differentiability condition as given  e.g.  in \cite{grun2019extending} and \cite{bakar2015modeling},
thus obtaining the closed-form estimates of the parameters.

The density function of composite GBII distribution that can be defined by matching two GBII distributions at 
an unknown threshold $u$ 
with unfixed mixing weight $r$ now can be written as
\begin{equation}
	\label{pdf: composite model}
	f_{\text{comGBII}}(y;\bm{\theta})=\begin{cases}
r  \displaystyle \frac{f_{\text{GBII}}(y;p_1,\mu_1,\nu_1,\tau_1)}{F_\text{GBII}(u;p_1,\mu_1,\nu_1,\tau_1)},& 0<y\leqslant u\\
		\\
		(1-r) \displaystyle \frac{f_{\text{GBII}}(y;p_2,\mu_2,\nu_2,\tau_2)}{1-F_\text{GBII}(u;p_2,\mu_2,\nu_2,\tau_2)},
		& u <y<\infty
	\end{cases}
\end{equation}
with $0\leqslant r\leqslant 1$,  
$\bm{\theta}=(p_1,\mu_1,\nu_1,\tau_1, p_2,\mu_2,\nu_2,\tau_2, r, u)$ denoting the vector of parameters in the composite GBII distribution, and
$p_1,\mu_1,\nu_1,\tau_1$ and $p_2,\mu_2,\nu_2,\tau_2$ denoting the vector of parameters in the head and the tail respectively.

The mode-matching procedure
is used by replacing the threshold $u$ by the modal value to satisfy the continuity and differentiability condition as the derivative at the mode is zero for unimodal distribution, yielding
\begin{align}
u&=y_{0}^1=y_{0}^2, \label{condition1}\\
	r  \displaystyle \frac{f_{\text{GBII}}(u;p_1,\mu_1,\nu_1,\tau_1)}{F_\text{GBII}(u;p_1,\mu_1,\nu_1,\tau_1)}&=(1-r) \displaystyle \frac{f_{\text{GBII}}(u;p_2,\mu_2,\nu_2,\tau_2)}{1-F_\text{GBII}(u;p_2,\mu_2,\nu_2,\tau_2)},
	\label{condition2}
\end{align}
where $y_{0}^1$ and $y_{0}^2$ denote the mode of the distributions used by the first and the second components of the composite model respectively.
Thus, the shape parameter $\mu_1$,
the threshold (the mode) $u$ and the mixing weight $r$
can be calculated based on the following equations:
\begin{align}
\label{eq: constrains1}
	\mu_1 & = \mu_2\left[\frac{p_2\nu_2-1}{p_2\tau_2+1}\right]^{1/p_2}\left[\frac{p_1\tau_1+1}{p_1\nu_1-1}\right]^{1/p_1},\\
	\label{eq: constrains2}
	u&=\mu_1 \left[\frac{p_1\nu_1-1}{p_1\tau_1+1}\right]^{1/p_1}=\mu_2 \left[\frac{p_2\nu_2-1}{p_2\tau_2+1}\right]^{1/p_2},\\
		\label{eq: constrains3}
	r&=\frac{I_{\nu_1,\tau_1}\left[\frac{p_1\nu_1-1}{p_1\nu_1+p_1\tau_1}\right]}{I_{\nu_1,\tau_1}\left[\frac{p_1\nu_1-1}{p_1\nu_1+p_1\tau_1}\right] + \phi \left\{1 - I_{\nu_2,\tau_2}\left[\frac{p_2\nu_2-1}{p_2\nu_2+p_2\tau_2}\right]\right\}},
\end{align}
where 
\begin{align}
 \phi &= \frac{p_1}{p_2}\frac{{B(\nu_2,\tau_2)}}{{B(\nu_1,\tau_1)}}
 \frac{(p_2\nu_2 + p_2\tau_2)^{(\nu_2 + \tau_2)}}{(p_1\nu_1 + p_1\tau_1)^{(\nu_1 + \tau_1)}	}
 \frac{(p_1\nu_1-1)^{\nu_1}(p_1\tau_1 + 1)^{\tau_1}}{(p_2\nu_2-1)^{\nu_2}(p_2\tau_2 + 1)^{\tau_2}}.
\end{align}

Note that the mixing weight $r$ only relies on the shape parameters $(p_1, p_2, \nu_1, \nu_2,\tau_1,\tau_2)$ and the threshold $u$ 
not only relies on these parameters but also the parameter $\mu_2$.
 It remains seven parameters 
$(\mu_2, p_1, p_2,  \tau_1, \tau_2, \nu_1, \nu_2)$,
all greater than 0, to be estimated with the two constrains $p_1\nu_1>1$ and $p_2\nu_2>1$
(the modes of the two GBII distributions must exist),
while three parameters $(\mu_1, u, r)$ are implicitly determined.

The cdf of composite GBII distribution is given by
\begin{equation}
	\label{cdf: composite-GBII}
F_{\text{comGBII}}(y;\bm{\theta})=\\
\begin{cases}
		r  \displaystyle \frac{F_\text{GBII}(y;p_1,\mu_1,\nu_1,\tau_1)}{F_\text{GBII}(u;p_1,\mu_1,\nu_1,\tau_1)},\quad 0 < y \le u\\ 
		\\
		r+(1-r) \displaystyle 
			\frac{F_\text{GBII}(y;p_2,\mu_2,\nu_2,\tau_2)-F_\text{GBII}(u;p_2,\mu_2,\nu_2,\tau_2)}{1-F_\text{GBII}(u;p_2,\mu_2,\nu_2,\tau_2)}
		,& u<y<\infty.
	\end{cases} 
\end{equation}

\begin{figure}[tp!]
	\centering
	\includegraphics[scale=0.6]{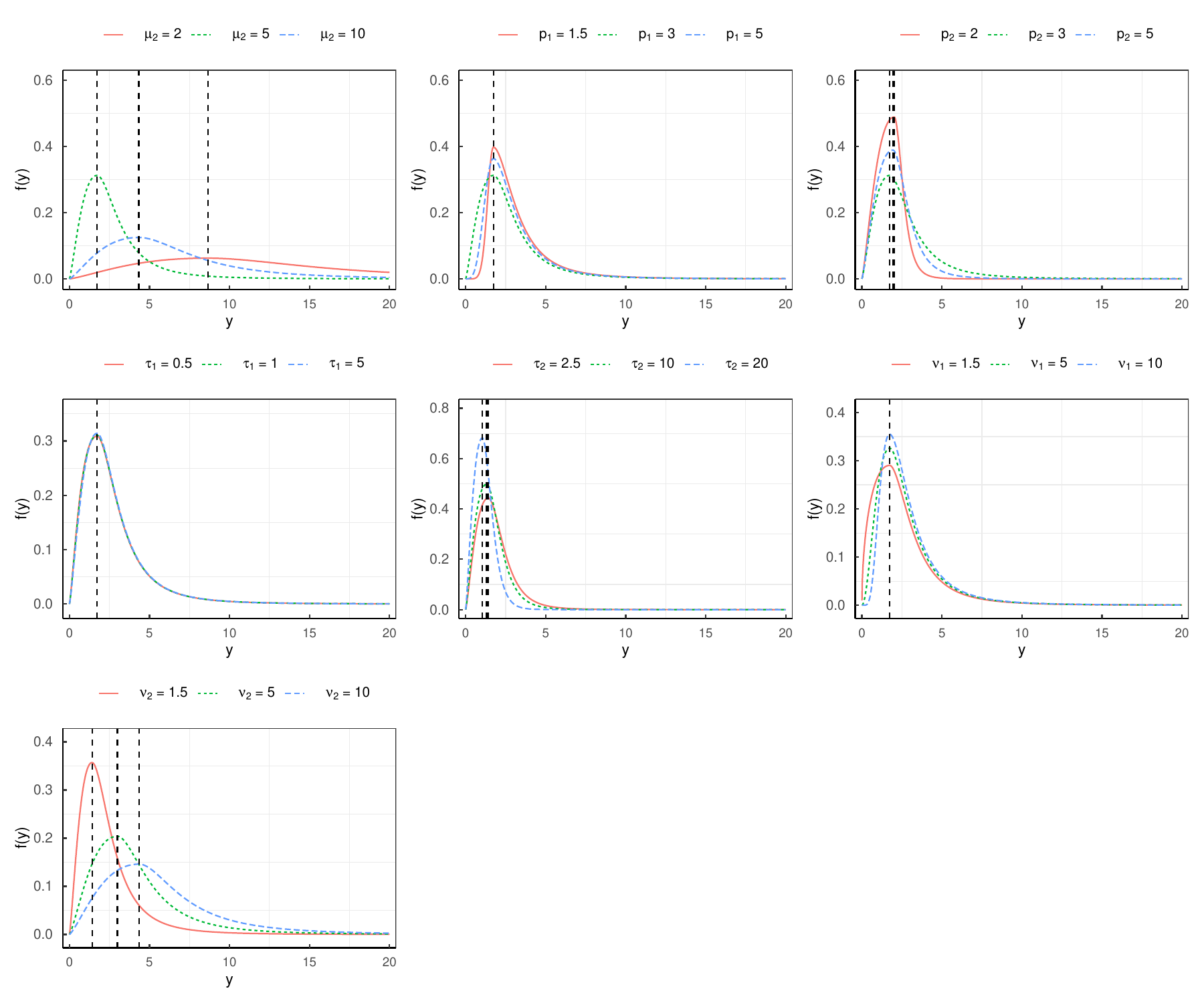}	
	\caption{{Density function for different composite GBII distributions
		when one parameter varies, keeping other
		parameters fixed.
		The default parameters are as followed: $\mu_2=2, p_1=1.5, p_2=2, \tau_1=2.5, \tau_2=1.5, \nu_1=1.5, \nu_2=2$.
		The thresholds are indicated by vertical dotted lines. }}
	\label{fig: pdf}
\end{figure}

To show how the parameters affect the shape of the composite GBII
distribution, 
Figure \ref{fig: pdf} plots the probability density functions when one parameter varies, keeping other
parameters fixed. 
The thresholds are indicated by vertical dotted lines. 
The plots show that in all cases, that the model has positive skewness.
The thresholds keep the same when the shape parameters $p_1, \tau_1, \nu_1$ in the first component varies,
and vary across the values of the parameters $\mu_2, p_2, \tau_2, \nu_2$ in the second component.
{
We also observe that the shape parameter $\tau_1$ does affect the density function when it has a small value, but its effect is relatively small compared to other parameters especially when it has a large value.  
It can lead to large variations when estimating the parameter $\tau_1$ with rather large standard errors.
Therefore, we need to re-parameterize the model when performing parameter estimation, see more details in Section \ref{sec: regression}.}

The $h$-th moments of the composite GBII distribution exist when $-p_1\nu_1<h<p_1\tau_1$ and $-p_2\nu_2<h<p_2\tau_2$:
\begin{align}
		\mathbb{E}\left[Y^h;\bm{\theta}\right]&=r{\mathbb{E}\left[Y^h|Y\le u\right]} + (1 -r)\mathbb{E}\left[Y^h|Y> u\right] , \nonumber
		\\
		&=r\mu_1^h\frac{B(\nu_1+h/p_1,\tau_1-h/p_1)}{B(\nu_1,\tau_1)}\frac{I_{\nu_1+h/p_1,\tau_1-h/p_1}\left[\frac{p_1\nu_1-1}{p_1\nu_1+p_1\tau_1}\right]}{I_{\nu_1,\tau_1}\left[\frac{p_1\nu_1-1}{p_1\nu_1+p_1\tau_1}\right]}\\
		&+(1-r)\mu_2^h\frac{B(\nu_2+h/p_2,\tau_2-h/p_2)}{B(\nu_2,\tau_2)}\frac{1-I_{\nu_2+h/p_2,\tau_2-h/p_2}\left[\frac{p_2\nu_2-1}{p_2\nu_2+p_2\tau_2}\right]}{1-I_{\nu_2,\tau_2}\left[\frac{p_2\nu_2-1}{p_2\nu_2+p_2\tau_2}\right]}.
		\nonumber
\end{align}

Next, a procedure for generating random variates from the composite GBII distribution is presented by using the inverse transformation method of simulation from the cdf of the two GBII distributions respectively.

\begin{itemize}
	\item let $q$ is generated from uniform distribution $q\sim \text{U}[0,1]$.
	\item if $q \leqslant r$, then the random samples is generated by using the following
	\begin{equation}
		\label{eq: VaR1}
		y=\mu_1\left[\frac{I_{\nu_1,\tau_1}^{-1}(z_1)}{1- I_{\nu_1,\tau_1}^{-1}(z_1)}\right]^{\frac{1}{p_1}},
	\end{equation}
	where $z_1={q}I_{\nu_1,\tau_1}\left[\frac{p_1\nu_1-1}{p_1\nu_1+p_1\tau_1}\right]/{r}$.
	\item if $q>r$, then
	\begin{equation}
		\label{eq: VaR2}
		y= \mu_2\left[\frac{I_{\nu_2,\tau_2}^{-1}(z_2)}{1- I_{\nu_2,\tau_2}^{-1}(z_2)}\right]^{\frac{1}{p_2}},
	\end{equation}
	where $z_2=F_{\text{GBII}}(u;p_2,\mu_2,\nu_2,\tau_2)+(q-r)(1-F_{\text{GBII}}(u;p_2,\mu_2,\nu_2,\tau_2))/(1-r)$.
\end{itemize}

We end this section computing some important risk measure expressions
for the composite GBII distribution. The Value-at-Risk (VaR)
with the quantile level $q$ was already given in \eqref{eq: VaR1} and \eqref{eq: VaR2}:
\begin{equation}
	\text{VaR}_{q}(Y;\bm{\theta})=\begin{cases}
	&\displaystyle \mu_1\left[\frac{I_{\nu_1,\tau_1}^{-1}(z_1)}{1- I_{\nu_1,\tau_1}^{-1}(z_1)}\right]^{\frac{1}{p_1}}, \quad \quad q \le r\\
	& \displaystyle \mu_2\left[\frac{I_{\nu_2,\tau_2}^{-1}(z_2)}{1- I_{\nu_2,\tau_2}^{-1}(z_2)}\right]^{\frac{1}{p_2}}, \quad \quad q > r.
	\end{cases}
\end{equation}

A closed-form expression for the tail Value-at-Risk (TVaR) can be obtained
as in \eqref{eq: incomplete-moments-2} with $s = \text{VaR}_q(Y;\bm{\theta})$ and $k = 1$:

\begin{equation}
	\text{TVaR}_{q}(Y;\bm{\theta})=\begin{cases}
		&\displaystyle 
		\mu_1\frac{B(\nu_1+1/p_1,\tau_1-1/p_1)}{B(\nu_1,\tau_1)}\frac{1-I_{\nu_1+1/p_1,\tau_1/p_1}(I_{\nu_1,\tau_1}^{-1}(z_1))}{1-z_1},
		\quad \quad q \le r\\
		& \displaystyle 
			\mu_2\frac{B(\nu_2+1/p_2,\tau_2-1/p_2)}{B(\nu_2,\tau_2)}
			\frac{1-I_{\nu_2+1/p_2,\tau_2-1/p_2}(I_{\nu_2,\tau_2}^{-1}(1-z_2)
				)}{1-z_2},
		\quad \quad q > r.
	\end{cases}
\end{equation}

\section{A composite GBII regression and statistical inference}\label{sec: regression}
\subsection{Regression modelling}
It is well known that the dependence of the response variable on the covariate(s) is modeled via the conditional mean of the response
variable in generalized linear models (GLMs). However, for heavy-tailed distributions the mean may not always exist.
Such distributions are extended to regression models with a link function between other model parameters such as the location, the scale or shape parameters and the covariates \citep{klein2014nonlife}. 
For instance, \cite{beirlant1998burr} discussed the regression modelling assuming the response variable follows the Burr distribution.
\cite{bhati2018generalized} assumed the response variable follows the generalized log-Moyal distribution with the shape parameter being a function of  the  covariates.  
\cite{li2021generalizing} constructed a new subclass models of GBII regressions.

{
	In the context of the composite GBII model, parameter identifiability is a crucial issue that needs to be carefully considered before incorporating covariates in a regression setting. The composite GBII distribution is susceptible to the problem of parameter identifiability due to the mathematical expression of the cdf in \eqref{cdf:GBII}, which states that $\text{GBII}(y;-p,\mu,\tau,\nu) = \text{GBII}(y;p,\mu,\nu, \tau)$. To address this, we assume that $p>0$ to avoid the unidentifiability problem of $\tau$ and $\nu$ in statistical inference. 
	In addition, we note that the three-parameter log-T (LT) distribution is a limiting case of the GBII distribution when the shape parameter $p$ approaches $0^+$,
	and the three-parameter generalized gamma (GG) is a limiting case when the shape parameter $\tau$ approaches $+\infty$. This presents difficulties in estimating the four parameters when $p$ has a very small estimated value or $\tau$ has a large value. 
Hence, we narrow our focus to the special cases of the GBII distribution rather than its limiting cases, where the body and the tail are constructed using specific values of the shape parameters, such as the Beta distribution of the second kind (BII), Burr (B) distribution, and others (see more details in Figure \ref{fig: gb2}).
	Moreover, we note that the GBII density is regularly varying at infinity with index $-p\tau-1$ and regularly varying at the origin with index $-p\nu-1$. This indicates that $p_2\nu_2$ controls the tail behavior of the proposed composite distribution, while $p_1\tau_1$ controls the body. Our preliminary study found that the estimated parameters, especially $p_1$, $\tau_1$ and $\nu_1$, are highly volatile, resulting in large standard errors. 
	To improve the stability of parameter estimation, 
	we assume that the response variable $Y_i$ follows the re-parameterized composite GBII distribution with $\sigma_1=p_1\tau_1, \sigma_2=p_2\tau_2, \delta_1 = p_1\nu_1, \delta_2=p_2\nu_2$, that is:
	$$Y_i \sim \text{Composite GBII}(\mu_2, p_1, p_2, \sigma_1, \sigma_2, \delta_1, \delta_2).$$}
We further propose the scale parameter $\mu_2(\bm{x}_i;\bm{\beta})$ to be modeled as a function of the
covariates $\bm{x}_i$ with regression coefficients $\bm{\beta}$ for $i$-th observation, $i=1,\dots, n$. 
In order to avoid boundary problems in optimization,
we consider a log-link function obtaining real values 
for the  shape parameters:
\begin{align}
	\label{eq: modelling setting}
	Y_i|\bm{x}_i &\sim \text{Composite GBII}(\bm{\beta}, \bm{\alpha}), \\
	\log \mu_2(\bm{x}_i;\bm{\beta})&=\bm{x}_i^T\bm{\beta},   \nonumber \\
	\log p_1 &= \alpha_1, \quad \log p_2 = \alpha_2,  \nonumber  \\
	\log \sigma_1 &= \alpha_3, \quad \log \sigma_2 = \alpha_4,  \nonumber  \\
	\log \delta_1 &= \alpha_5, \quad \log \delta_2 = \alpha_6,  \nonumber 
\end{align}
where $\bm{\alpha}=(\alpha_1,...,\alpha_6)$ denotes the vector of log transformation of model's parameters,
$\bm{x}_{i}^T=(1,x_{i1},...,x_{ik})$ denotes the vector of covariates,
$\bm{\beta}=(\beta_0,\beta_1,...,\beta_k)^T$ the vector of coefficients.
{
In order to enhance the interpretability of the model parameters, we refer to the aforementioned parameters as intermediate parameters  solely used in the estimation process. We report the shape parameters' estimation results without re-parameterization, obtained as $\tau_1=\sigma_1/p_1, \tau_2 = \sigma_2/p_2, \nu_1=\delta_1/p_1$, and $\nu_2=\delta_2/p_2$.
It enables us to explicitly identify the key shape parameters and facilitates the interpretation of the results obtained from statistical inference, especially when some subclass of GBII family are used in the simulation and application study.
}

The $h$-th moment of the composite GBII distribution depends on the $(k+1)$-dimensional vector of
covariates $\bm{x}_{i}$
which is given by
\begin{equation}
\label{eq: moment-regression}
\mathbb{E}\left[Y^h_i; \bm{\beta}, \bm{\alpha}\right]=w(\bm{\alpha})\exp(\bm{x}_i^T\bm{\beta})^h,
\end{equation}
where
\begin{align*}
w(\bm{\alpha}):&=r\frac{g_2^{1/p_2}}{g_1^{1/p_1}}\frac{B(\nu_1+h/p_1,\tau_1-h/p_1)}{B(\nu_1,\tau_1)}\frac{I_{\nu_1+h/p_1,\tau_1-h/p_1}\left[\frac{p_1\nu_1-1}{p_1\nu_1+p_1\tau_1}\right]}{I_{\nu_1,\tau_1}\left[\frac{p_1\nu_1-1}{p_1\nu_1+p_1\tau_1}\right]}\\
&+(1-r)\frac{B(\nu_2+h/p_2,\tau_2-h/p_2)}{B(\nu_2,\tau_2)}\frac{1-I_{\nu_2+h/p_2,\tau_2-h/p_2}\left[\frac{p_2\nu_2-1}{p_2\nu_2+p_2\tau_2}\right]}{1-I_{\nu_2,\tau_2}\left[\frac{p_2\nu_2-1}{p_2\nu_2+p_2\tau_2}\right]}.
\end{align*}
Note that  $w(\bm{\alpha})$ only depends on the  shape parameters which are not related to covariates.
Similar to the conventional GLMs, it is obvious that the mean of the composite GBII regression model are proportional to some exponential transformation of the linear predictor $\bm{x}_i^T\bm{\beta}$, which can provide an intuitive interpretation for the insurance pricing and reserving. 
Also, it provides the explicit expressions of VaR across all individuals and related to covariates when the mean does not always exist for modelling more extreme losses.
Moreover, 
the regression setting in \eqref{eq: modelling setting} also allows a varying threshold across policyholders based on their observed risk features:
\begin{equation}
\label{eq: varying-threshold}
u(\bm{x}_i; \bm{\alpha}, \bm{\beta})= \mu_2(\bm{x}_i;\bm{\beta})\left[\frac{p_2\nu_2-1}{p_2\tau_2+1}\right]^{1/p_2}.
\end{equation}
In such cases, the policyholder heterogeneity among the individuals in the tail parts of loss data can be captured sufficiently by the proposed regression model.

\subsection{Constrained maximum likelihood estimation method }
In this section, we will show how to perform the constrained maximum likelihood (CML) estimation method to obtain the values of the parameter vector $\bm{\beta}$ and $\bm{\alpha}$.
Given a data set $\bm{y}=(y_1,...,y_n)$, the log-likelihood function of the proposed composite GBII regression is given by 
{{
		\begin{align*}
		\ell(\boldsymbol{\beta}, \boldsymbol{\alpha} ; \boldsymbol{y})= & \sum_{i=1}^n \log [r(\boldsymbol{\alpha})] I\left[y_i \leq u\left(\boldsymbol{x}_i ; \boldsymbol{\beta}\right)\right]+\sum_{i=1}^n \log f_{\mathrm{GBII}}\left(y_i ; \mu_1\left(\boldsymbol{x}_i ; \boldsymbol{\beta}\right),  \sigma_1, \sigma_2, \delta_1, \delta_2\right) I\left[y_i \leq u\left(\boldsymbol{x}_i ; \boldsymbol{\beta}\right)\right] \\
		- & \sum_{i=1}^n \log I_{\delta_1/p_1, \sigma_1/p_1}\left[\frac{\delta_1-1}{\delta_1+\sigma_1}\right] I\left[y_i \leq u\left(\boldsymbol{x}_i ; \boldsymbol{\beta}\right)\right] \\
		& +\sum_{i=1}^n \log [1-r(\boldsymbol{\alpha})] I\left[y_i>u\left(\boldsymbol{x}_i ; \boldsymbol{\beta}\right)\right]+\sum_{i=1}^n \log f_{\mathrm{GBII}}\left(y_i ; \mu_2\left(\boldsymbol{x}_i ; \boldsymbol{\beta}\right), \sigma_1, \sigma_2, \delta_1, \delta_2\right) I\left[y_i>u\left(\boldsymbol{x}_i ; \boldsymbol{\beta}\right)\right] \\
		& -\sum_{i=1}^n \log \left\{1-I_{\delta_2/p_2, \sigma_2/p_2}\left[\frac{ \delta_2-1}{ \delta_2+ \sigma_2}\right]\right\} I\left[y_i>u\left(\boldsymbol{x}_i ; \boldsymbol{\beta}\right)\right] .
		\end{align*}}}

{The proposed CML estimation can be accomplished relatively easily subjected to 
$\delta_1>1$ and $\delta_2>1$ by solving the following nonlinear programming problem:}
\begin{align}
\label{eq: constrains-0}
&\max_{\bm{\beta}, \bm{\alpha}\in \mathbb{R}^{k+7}} \ell(\bm{\beta}, \bm{\alpha};\bm{y})  \\
\text{s.t.}&
\quad \quad h_1(\bm{\alpha})=\epsilon_1-\alpha_5 \le 0, \quad h_2(\bm{\alpha})=\epsilon_2-\alpha_6\le 0, \nonumber 
\end{align}
where $k$ is the number of covaraites, $\epsilon_1$ and $\epsilon_2$ denote the small enough values with positive support, e.g, $\epsilon_1=\epsilon_2=1.0e^{-6}$.
The values of the parameter vector $\bm{\beta}$ and $\bm{\alpha}$ can be obtained using augmented Lagrange multiplier method (see e.g. \cite{nocedal2006numerical})  which is 
a class of algorithms
for constrained nonlinear optimization that enjoy favorable theoretical properties for finding
local solutions from arbitrary starting points.
Thus, the objective function for the inequality constrained problem \eqref{eq: constrains-0} is given by
\[
 L_{\rho}(\bm{\beta},\bm{\alpha},\bm{\lambda}):=-\ell(\bm{\beta},\bm{\alpha};\bm{y}) +\sum_{k=1}^{2}\lambda_k h_k(\bm{\alpha}) + \sum_{k=1}^{2}\frac{\rho}{2}\left[\max(0, h_k(\bm{\alpha}))^2  \right],
\]
where $\bm{\lambda}=(\lambda_1, \lambda_2)\in \mathbb{R}_{+}^{2}$ are the Lagrange multipliers and $\rho>0$ is a penalty parameter.
Algorithm \ref{alg:ALMGB} gives the details of the augmented Lagrange multiplier method for CML estimation.
The 
gradients 
of the log-likelihood function $\ell(\bm{\beta}, \bm{\alpha};\bm{y})$ are needed in this step, which are given in Appendix \ref{app:Gradients}.
The asymptotic variance-covariance matrix can be computed as the inverse of the observed Fisher information matrix, which can be obtained using the second-order derivatives of the log-likelihood function $\ell(\bm{\beta}, \bm{\alpha};\bm{y})$.
{The standard errors for the shape parameters $p_1, p_2, \nu_1, \nu_2, \tau_1, \tau_2$ can be obtained by the delta method \citep{oehlert1992note}.}
The estimation results are obtained using 
constrained optimization by sequential quadratic programming (SQP) optimization numerical optimization 
via 
function \texttt{solnp} as part of the package \textbf{Rsolnp} in R software.
For additional information regarding this optimization
procedure, we refer the reader to \cite{ye1987interior} and \cite{gill2005snopt}.

\begin{algorithm}
	\caption{augmented Lagrange multiplier method for composite GBII regression}
	\label{alg:ALMGB}
	\KwData{positive data points $\bm{y}=(y_1,...,y_n)$, covariates $\bm{x}_{1},...,\bm{x}_n$, and initial parameters $\bm{\beta}^{(0)}$ and $\bm{\alpha}^{(0)}$, Lagrange multiplier vector $\bm{\lambda}^{(0)}$, 
		penalty parameter $\rho^{(0)}$, increment $c$, tolerance $\epsilon$;}
 \KwResult{CML estimaties $\bm{\beta}^{*}, \bm{\alpha}^{*}$ and $\bm{\lambda}^{*}$;}
 set $t=0$\;
	\While{$\lVert\nabla_{\bm{\beta}, \bm{\alpha}}L_{\rho^{(t)}}(\bm{\beta}^{(t)},\bm{\alpha}^{(t)}, \bm{\lambda}^{(t)}) \lVert /\Vert L_{\rho^{(t)}}(\bm{\beta}^{(t)},\bm{\alpha}^{(t)}, \bm{\lambda}^{(t-1)})\lVert\ge \epsilon$}{	
		solve for $\bm{\beta}^{(t+1)}= \arg \min_{\bm{\beta}} L_{\rho^{(t)}}(\bm{\beta}^{(t)},\bm{\alpha}^{(t)}, \bm{\lambda}^{(t)})$\;
		solve for $\bm{\alpha}^{(t+1)}= \arg \min_{ \bm{\alpha}} L_{\rho^{(t)}}(\bm{\beta}^{(t)},\bm{\alpha}^{(t)}, \bm{\lambda}^{(t)})$\;
		update the Lagrange multipliers: ${\lambda}_j^{(t+1)}=  {\lambda}_j^{(t)} + 2\rho^{(t)} h_j(\bm{\alpha}^{*(t)})$, $j=1,2$ \;
		set $\rho^{(t+1)}= c \rho^{(t)}$ and $t = t + 1$\;
}

\end{algorithm}

\subsection{Computational details}

{
(1) Initialization of $\bm{\beta}$ and $\bm{\alpha}$.
To ensure accurate estimation results using Algorithm \ref{alg:ALMGB}, proper initialization of both $\bm{\beta}$ and $\bm{\alpha}$ is crucial, especially for the highly sensitive log-transformed parameters $\bm{\alpha}$. As this algorithm may only converge to a local minimum for problems with rugged landscapes, it is important to address the issue of local optima. To achieve this, we utilize the \texttt{glm} function in R to fit a univariate Gamma regression model using a log-link function, and employ the resulting regression coefficients as the initial values $\bm{\beta}^{(0)}$ for the optimization process. For the remaining parameters, we utilize a random initialization strategy, drawing the log-transformed parameters from a standard normal distribution while taking into account the parameter constraints outlined in \eqref{eq: modelling setting}.
The final initialization parameters $\bm{\alpha}^{(0)}$ are selected with the smallest negative log-likelihood value from multiple sets of random generation. 
The simulation study in Section \ref{sec: simulation} shows this initialization strategy enables us to obtain a reasonable starting point for the optimization process, resulting in the achievement of accurate estimation results.
}


(2) Model selection. To compare models with the different number of parameters, in terms of goodness-of-fit, 
we consider the Akaike
information criterion (AIC) and the Bayesian information criterion (BIC), defined respectively as 
\begin{align*}
	\text{AIC}&=-2\ell(\bm{\beta},\bm{\alpha};\bm{y})+2m,\\
	\text{BIC}&=-2\ell(\bm{\beta},\bm{\alpha};\bm{y})+m\log n,
\end{align*}
where $\ell$ denotes the log-likelihood value, $m$ the number of model parameters and $n$ the number of observations.
The BIC gives more penalties than AIC does.
The model with the minimum AIC or BIC value is selected as the preferred model to fit the data whereas the BIC gives more penalties than AIC.


For assessment of the regression model we also use  randomized quantile residuals  defined by
$r_{i}=\Phi^{-1}\left[F_{\text{comGBII}}(y_{i};\hat{\bm{\beta}},\hat{\bm{\alpha}})\right]$ ($i=1,\ldots,n$),
where $\Phi\left(\cdot\right)$ is the cdf of the standard normal distribution and $F_{\text{comGBII}}$ denotes the cdf of the composite GBII model as given in \eqref{cdf: composite-GBII}. 
The distribution of $r_{i}$ converges to standard normal if $(\hat{\bm{\beta}},\hat{\bm{\alpha}})$ are consistently estimated, see \cite{dunn1996randomized}, and hence a normal QQ-plot of the $r_i$ should follow the 45 degree line for the regression application to be relevant.



%
%

\section{Simulation study}\label{sec: simulation}

In this section, we first check the accuracy of the CML
estimators 
discussed in
Section \ref{sec: regression},
and then evaluate the performance of  the proposed model, compared with the conventional GLMs 
in the case that the tail behavior of the simulated loss data is
explained by several observed covariates which characterize the individuals.

\begin{figure}[tp!]
	\centering
	\includegraphics[scale=0.7]{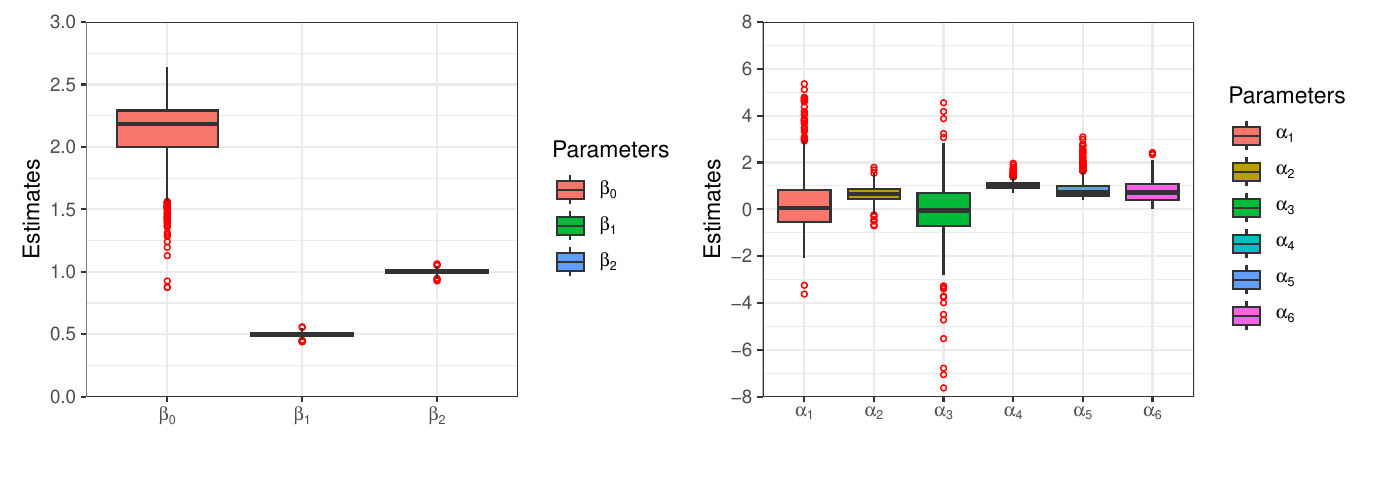}	
	\caption{{Boxplots of the parameter estimates from 1000 composite GBII simulated samples of size $n=2000$. \textit{Left}: results for the regression coefficients $\beta_0, \beta_1$ and $\beta_2$.
		\textit{Right}: results for the log-transformation of model's parameters $\alpha_1, \alpha_2, \alpha_3, \alpha_4, \alpha_5, \alpha_6$. }
	}
	\label{fig: simulation-boxplot-revised}
\end{figure}

{
We generated $N = 1,000$ data sets of sample size $n=2,000$ from the composite GBII model with 
$k=2$, $\bm{\beta}=(2, 0.5, 1)$, $\bm{\alpha}=(\log(1), \log(1.5), \log(1.5), \log(3), \log(2), \log(3))$,
$\bm{x}_i^T=(1, x_{i1},x_{i2})$ and the covariates $x_{i1}$ and $x_{i2}$ being generated from the standard normal distribution.
Thus, the true parameters of the composite GBII regression are set as: $\mu_{2}(\bm{x}_{i})=\exp(2 + 0.5 x_{i1} + 1x_{i2})$,
$p_1 = 1, p_2=1.5, \tau_1 = 1.5, \tau_2 = 2, \nu_1 = 2$ and $\nu_2=2$.
The maximum likelihood estimates for each simulation are calculated by using CML method.
In Figure \ref{fig: simulation-boxplot-revised}, we present the boxplots of the parameter estimates $\bm{\beta}$ and $\bm{\alpha}$ from 1000
Monte Carlo simulations.
The median estimate for $\beta_0$ is 2.18, with a mean of 2.11. The estimates for $\beta_1$ and $\beta_2$ are very close to the true values with little variation.
The estimates for $\alpha_1, \alpha_2, \alpha_4, \alpha_5,$ and $\alpha_6$ overestimate the true values, while the estimate for $\alpha_3$ underestimates the true value. 
Outliers are present in the sample estimates, especially for $\beta_0, \alpha_1,$ and $\alpha_3$.
The simulation results suggest that the proposed composite GBII model performs well in capturing the relationship between the covariates and the response variable. 
The model estimates for the parameters $\bm{\alpha}$ show some bias, but overall, the model provides reasonable estimates for both the shape and scale parameters. 
}

To further investigate the global fitting performance of the proposed composite GBII regression model  
when both the body and the tail behavior of the loss data vary across individuals that are related to several covariates,
we simulate the $n=2000$ samples from the following steps:
\begin{itemize}
	\item 
	The covariates $x_{i1}$ and $x_{i2}$ is generated from the standard normal distribution for $i=1,...2000$, and consider the loss data being simulated from two components: the body and the tail part receptively. 
	The covariates both have effects on small and large amounts.
	\item 
	The loss data in the first component is generated from a Gamma distribution with the mean parameter $\mu_{\text{GA}}=\exp(2 + 2x_{i1}+0.5x_{i2})$, the dispersion parameter $\phi_{\text{GA}}=1$, which is denoted by $Y_{i}\sim \text{GA}(\mu_{\text{GA}}(\bm{x}_i), \phi_{\text{GA}})$ for $i=1,...,1800$.
	\item 
	The loss data in the second component is generated from a generalized Pareto distribution (GPD) with the location parameter $\mu_{\text{GPD}}=\exp(5 + 0.5x_{i1}+0.5x_{i2})$, the scale parameter 
	$\sigma_{\text{GPD}}=\exp(3 + 0.5x_{i1}+1x_{i2})$
	and the shape parameter $\xi_{\text{GPD}}=1$, which is denoted by $Y_{i}\sim \text{GPD}(\mu_{\text{GPD}}(\bm{x}_i), \sigma_{\text{GPD}}(\bm{x}_i), \xi_{\text{GPD}})$
	for $i=1801,...,2000$. Note that the location parameter in the GPD is regarded as the threshold that is modelled as a function of  individual risk features in this regression setting.
\end{itemize}

\begin{figure}[tp!]
	\centering
	\includegraphics[scale=0.7]{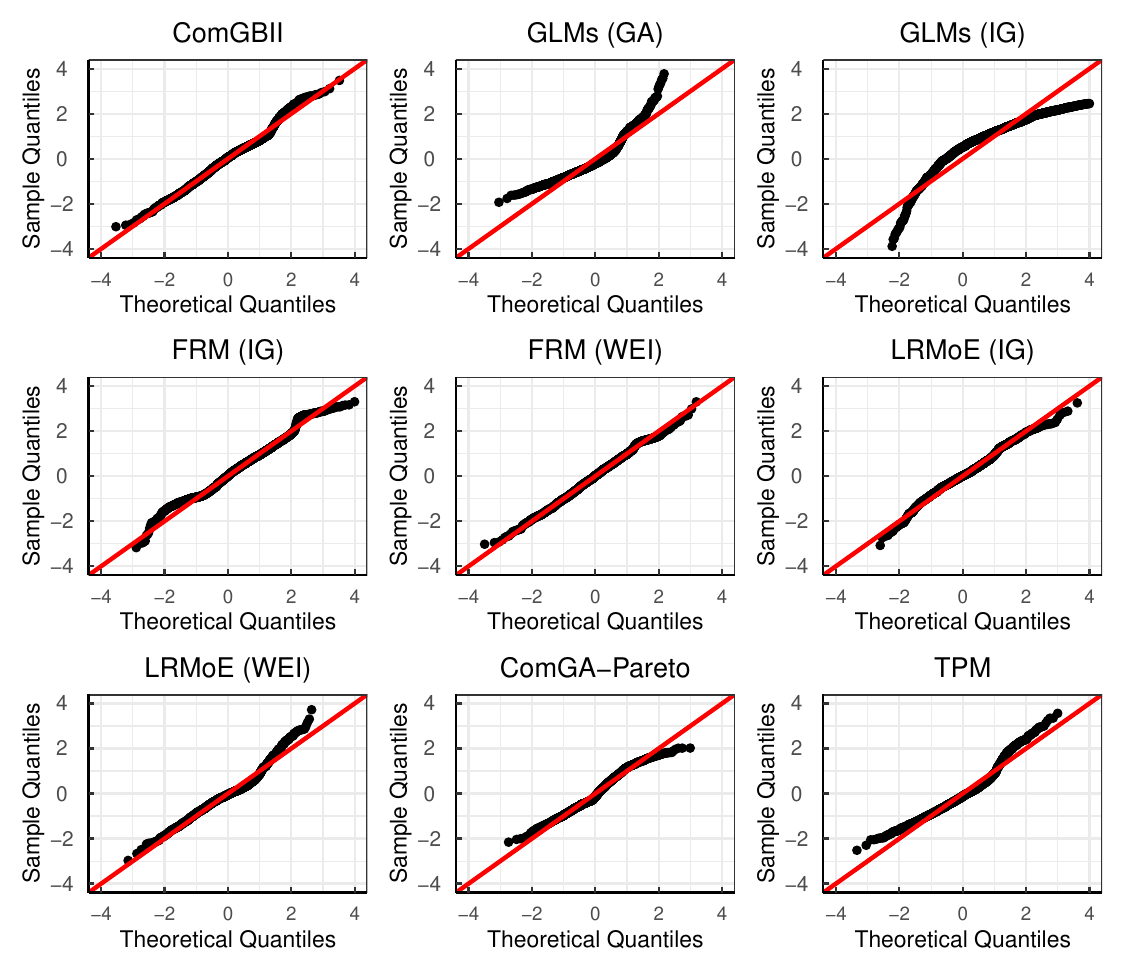}	
	\caption{\red{Normal QQ-plots of quantile residuals from the composite GBII regression and competing models in one Monte Carlo simulation with
		sample size $n=2000$.}}
	\label{fig: simulation-qq}
\end{figure}

We compare the performance of the proposed composite GBII model with several state-of-the-art models and two standard models on a simulated data set. The competing models include the finite mixture regression model (FMR), the mixture-of-experts regression model (MoE) 
\citep{fung2019class}, 
the two-part regression model (TPM), 
the composite Gamma-Pareto regression model \citep{gan2018fat}, as well as the standard GLM Gamma (GA) and GLM inverse Gaussian (IG) models.
Specifically, we consider the logit-weighted reduced mixture of experts model (LRMoE) for MoE model \citep{fung2022fitting}.
\red{
Given that the simulated data features heavy-tailed distributions, we use the Inverse Gaussian and Weibull distributions respectively as the mixture components in the mixture models for comparison.
The \textbf{gamlss.mx} package within R software is used to obtain the estimation results.
}
 In the TPM, the loss data below a given threshold is modelled by the truncated Gamma regression and the tail data is modelled by a generalized Pareto regression model \footnote{The scale parameter of the GPD is modelled by a exponentially linear combination of the covariates.}. 
 The threshold is pre-specified and the probability of exceeding the threshold is estimated using a GLM logit-model \citep{laudage2019severity}.
The composite Gamma-Pareto model can also be seen as a two-part regression model, where the threshold is an unknown parameter to be estimated. 
The TPM and the composite Gamma-Pareto are similar in that the covariates are introduced to model the body and the tail, respectively, with a non-varying threshold.
\red{
Table \ref{tab:estimates - simulation - 2} highlights the superior performance of the FMR (WEI) and composite GBII models over other models based on three key metrics: negative log-likelihood values (NLL), AIC, and BIC values. In addition, Figure \ref{fig: simulation-qq} displays normal QQ-plots of quantile residuals from these models. Notably, the TPM and composite Gamma-Pareto models exhibit subpar performance, primarily due to their assumption of a constant threshold across covariates, limiting their ability to accurately capture tail data. In contrast, the composite GBII model and FRM (WEI) effectively capture the distributional characteristics of the data, especially in cases of heavy-tailed distributions. They do so by accommodating varying thresholds and addressing different tail behaviors for all individuals based on their observed risk features.
}




\begin{table}
	\caption{\red{simulation: model selection measures.}}
	\begin{tabular*}{\hsize}{@{}@{\extracolsep{\fill}}llcccccc@{}}
		\toprule
    \multicolumn{2}{c}{Models} & Npars & NLL   & AIC   & Rnaking & BIC   & Ranking \\
    \hline
\multirow{3}[0]{*}{Composite models} & ComGBII & 9     & 7417.62 & 14853.25 & 2     & 14903.65 & 2 \\
& ComGA-Pareto & 11    & 8437.472 & 16896.94 & 8     & 16958.55 & 8 \\
& TPM   & 11    & 8135.99 & 16293.98 & 4     & 16355.59 & 4 \\
\hline
\multirow{4}[0]{*}{Mixture models} & FRM (IG) & 9     & 8326.08 & 16670.15 & 6     & 16720.56 & 6 \\
& LRMoE (IG) & 7     & 8149.08 & 16318.16 & 5     & 16357.37 & 5 \\
& FRM (WEI) & 9     & 7244.44 & 14506.89 & 1     & 14557.30 & 1 \\
& LRMoE (WEI) & 7     & 7970.49 & 15960.98 & 3     & 16000.19 & 3 \\
\hline
\multirow{2}[0]{*}{GLMs} & GLM (GA) & 4     & 8409.53 & 16827.06 & 7     & 16849.46 & 7 \\
& GLM (IG) & 4     & 10177.39 & 20362.77 & 9     & 20385.17 & 9 \\
		\bottomrule
	\end{tabular*}%
	\begin{tablenotes}
		\item \small 	Npars denotes the number of estimated parameters. 
		\item \small FMR  (IG) and FMR (WEI) denote the FMR model with two inverse Gaussian distributions and  two Weibull distribution as mixture components respectively. ComGA-Pareto denotes the composite Gamma-Pareto regression model.
	\end{tablenotes}
	\label{tab:estimates - simulation - 2}
\end{table}

\newpage

\section{Applications}\label{sec: applications}
In this section, we will illustrate the proposed method with the two practical
examples.
The first example concerns modelling the univariate loss data set without covariates by considering well-known danish fire insurance data.
The second example is about investigating the behavior of the proposed regression model with covariates for modelling the short-term medical insurance claim data from a Chinese insurance company.

\subsection{Danish fire insurance data set}\label{section:application1}
As the first example, we fit the univariate composite GBII distribution and its subfamilies to the well-known danish fire insurance data.
The data set contains of $n=2492$ fire losses, in millions of
Danish Krone (DKK) for the period 1980-1990 inclusively,  and have been adjusted to reflect to inflation.
The following reports some summary statistics for the data: 
minimum is 0.3134,
25\% quantile is 1.1570, 0.75\% quantile is 2.6450, maximum is 263.3, mean is 3.0630, and standard deviation is 7.9767.
This data set is available in the R package \textbf{SMPracticals}, which was also studied in 
\cite{cooray2005modeling},
\cite{scollnik2007composite},
\cite{calderin2016modeling},
\cite{scollnik2012modeling},
\cite{bakar2015modeling},
\cite{nadarajah2014}, and
\cite{grun2019extending}
using a variety of composite models.

We now specify several composite GBII models paying special attention to the subfamily of GBII distribution.
Several loss distributions that the GBII nests as special cases are considered (\cite{klugman2012loss}; p.669-681):
\begin{itemize}
	\item the three-parameter Burr (Singh-Maddala distribution) and inverse Burr are obtained for $\nu=1$ and $\tau=1$ respectively. 
	\item the three-parameter Beta distribution of the second kind is obtained for $p=1$.
	\item the three-parameter GLMGA is obtained for $\nu=\frac{1}{2}$.
	\item the two-parameter Paralogistic and inverse Paralogistic are obtained for $\tau=p,\nu=1$ and $\tau=1,\nu=p$ respectively. 
\end{itemize}

The details of distributions nested within the GBII distribution are shown in Appendix \ref{app:sub-GBII}.
The composite models we considered are restricted to those
with a three-parameter GLMGA (G) distribution forming the tail and the head belonging to the GBII and its subfamily consisting of two or three parameters:
the GBII,
the Beta distribution of the second kind (BII), the Burr (B), the inverse Burr (IB),
the Paralogistic (P) and the inverse Paralogistic (IP).
Thus, the seven composite models are given as follows \footnote{{The re-parameterized composite distributions are shown in Appendix \ref{app:re-parameterized-GBII}.}}:
\begin{itemize}
	\item Composite GBII model (ComGBII) $( \mu_2, p_1, p_2, \tau_1, \tau_2, \nu_1, \nu_2)$.
	\item GBIIG model $( \mu_2, p_1, p_2, \tau_1, \tau_2, \nu_1, \nu_2 = \frac{1}{2})$.
	\item BIIG model $( \mu_2, p_1=1, p_2, \tau_1, \tau_2, \nu_1, \nu_2 = \frac{1}{2})$.
	\item BG model $( \mu_2, p_1, p_2, \tau_1, \tau_2, \nu_1=1, \nu_2 = \frac{1}{2})$.
	\item IBG model $( \mu_2, p_1, p_2, \tau_1=1, \tau_2, \nu_1, \nu_2 = \frac{1}{2})$.
	\item PG model $(\mu_2, p_1, p_2, \tau_1=p_1, \tau_2, \nu_1=1, \nu_2=\frac{1}{2})$.
	\item IPG model $(\mu_2, p_1, p_2, \tau_1=1, \tau_2,   \nu_1=p_1 ,  \nu_2=\frac{1}{2})$.
\end{itemize}

In Table \ref{tab:estimates - danish - 1} we provide the estimates and standard errors of the seven models above.
The method of constrained maximum likelihood was used and the standard errors were computed by inverting the observed information matrix.
{One can see that the estimates of the scale parameter $\mu_2$ in these models are all around 1, while the estimated shape parameters are different across the models.
Estimated shape parameters tend to be stable with relatively small standard error estimates for most of models except BIIG model.
Table \ref{tab:estimates - danish - 2}
reports the NLL, the AIC and the BIC values of the proposed models.
The results of six composite models discussed in \cite{grun2019extending} are also reported in  Table \ref{tab:estimates - danish - 2} for model comparison. 
{
It is clear from Table \ref{tab:estimates - danish - 2} that the ComGBII model shows the smallest NLL value as it has the most parameters.}
The IBG model provides a best fit with the smallest AIC and BIC values, followed by the GBIIG (according to AIC) and PG model (according to BIC).}
While the WIW, PIW and IBW models proposed by \cite{grun2019extending} rank third to fifth in terms of BIC, with the Inverse Weibull being used in the tail part, the differences between these models are minimal as their BIC values are very close.

The goodness-of-fit measures and the bootstrap P-values
for the corresponding goodness-of-fit tests of all the competing models are reported in Table \ref{tab:gof - danish}. 
We consider the Kolmogorov-Smirnov (KS), Anderson-Darling (AD) and Cram\'er-von Mises (CvM) test statistics with corresponding P-values, choosing for the models with small values of the KS, AD and CvM test statistics, or large values of the corresponding P-values. The P-values are obtained using the bootstrap method as developed in \cite{calderin2016modeling}. 
Here again the proposed seven models are prevailing with a P-value above 0.6, which all shows the better fit than the competing models.

\begin{table}[hbt!]
	\centering
	\caption{{Danish fire insurance claims: model estimation.}}
	\begin{tabular*}{\hsize}{@{}@{\extracolsep{\fill}}cccccccc@{}}
    	\toprule
	\textbf{Models/Parameters} & $\mu_2$   & $p_1$    & $p_2$    & $\tau_1$  & $\tau_2$  & $\nu_1$   & $\nu_2$ \\
	\hline
    \multirow{2}[0]{*}{ComGBII} & 1.12  & 435.60 & 5.10  & 0.03  & 0.28  & 0.04  & 0.35 \\
& (0.00)   & (1.89) & (1.81) & (0.00)   & (0.36) & (0.00)   & (0.13) \\
\multirow{2}[0]{*}{GBIIG} & 1.02  & 89.09 & 5.04  & 0.004  & 0.28  & 0.16  & 0.50 \\
& (0.02) & (1.79) & (5.46) & (0.00)   & (3.46) & (0.00)   &  (-) \\
\multirow{2}[0]{*}{BIIG} & 1.07  & 1.00  & 5.73  & 20268.01 & 0.25  & 44.92 & 0.50 \\
& (0.00)   &  (-) & (0.02) & (47.01) & (1.14) & (0.81) &  (-) \\
\multirow{2}[0]{*}{BG} & 1.03  & 16.23 & 5.09  & 345.74 & 0.28  & 1.00  & 0.50 \\
& (0.00)   & (0.02) & (0.29) & (1.67) & (0.32) &  (-) &  (-) \\
\multirow{2}[0]{*}{IBG} & 1.04  & 447.17 & 4.50  & 1.00  & 0.32  & 0.04  & 0.50 \\
& (0.00)   & (1.48) & (1.78) &  (-) & (0.13) & (0.00)   &  (-) \\
\multirow{2}[0]{*}{PG} & 1.05  & 16.42 & 4.85  & 16.42 & 0.30  & 1.00  & 0.50 \\
& (0.01) & (0.21) & (0.38) & (0.21) & (0.02) &  (-) &  (-) \\
\multirow{2}[0]{*}{IPG} & 1.09  & 5.12  & 5.80  & 1.00  & 0.25  & 5.12  & 0.50 \\
& (0.00)   & (0.00)   & (0.40) &  (-) & (0.03) & (0.00)   &  (-) \\
	\bottomrule
\end{tabular*}%
	\label{tab:estimates - danish - 1}%
	\begin{tablenotes}
		\item \small The standard errors of estimates are reported in parentheses, which are calculated by using delta method.
	\end{tablenotes}
\end{table}

\begin{table}[hbt!]
		\caption{{Danish fire insurance claims: model selection measures.}}
				\begin{tabular*}{\hsize}{@{}@{\extracolsep{\fill}}lcccccc@{}}
\toprule
			Model & Npars & NLL   & AIC   & Ranking & BIC   & Ranking \\
			\hline
ComGBII & 7     & \textbf{3813.87} & 7641.74 & 4     & 7682.49 & 11 \\
GBIIG & 6     & 3813.99 & 7639.99 & 2     & 7674.91 & 8 \\
BIIG  & 5     & 3850.38 & 7710.76 & 12    & 7739.87 & 13 \\
BG    & 5     & 3817.92 & 7645.83 & 8     & 7674.94 & 10 \\
IBG   & 5     & 3814.02 & \textbf{7638.03} & 1     & \textbf{7667.13 }& 1 \\
PG    & 4     & 3818.32 & 7644.63 & 7     & 7667.92 & 2 \\
IPG   & 4     & 3853.58 & 7715.16 & 13    & 7738.45 & 12 \\
\hline
WIW   & 4     & 3820.01 & 7648.02 & 9     & 7671.30 & 3 \\
PIW   & 4     & 3820.14 & 7648.28 & 10    & 7671.56 & 4 \\
IBW   & 5     & 3816.34 & 7642.68 & 5     & 7671.79 & 5 \\
WIP   & 4     & 3820.93 & 7649.87 & 11    & 7673.15 & 6 \\
IBP   & 5     & 3817.07 & 7644.14 & 6     & 7673.25 & 7 \\
IBB   & 6     & 3814.00 & 7639.99 & 3     & 7674.92 & 9 \\
			\bottomrule
		\end{tabular*}%
		\begin{tablenotes}
		\item \small  WIW, PIW, IBW, WIP, IBP, IBB represent 
		Weibull-Inverse Weibull,
		Paralogistic-Inverse Weibull,
		Inverse Burr-Inverse Weibull,
		Weibull-Inverse Paralogistic,
		Inverse Burr-Inverse Paralogistic and
		Inverse Burr-Burr composite distribution proposed in \cite{grun2019extending}.	
		The six models are selected as the 6 best fitting composite models (according to the BIC) when estimated to the Danish fire loss data.
	\end{tablenotes}
				\label{tab:estimates - danish - 2} 
\end{table}

\begin{table}[hbt!]
	\centering
	\caption{{Danish fire insurance claims: correlation \textit{R} of QQ-plots, and KS, AD and CvM goodness-of-fit tests.}}
	\begin{tabular*}{\hsize}{@{}@{\extracolsep{\fill}}llrlrlrl@{}}
		\bottomrule
		\multirow{2}[4]{*}{Model} &  \multirow{2}[4]{*}{R}& \multicolumn{2}{l}{Kolmogorov-Smirnov} & \multicolumn{2}{l}{Anderson-Darling} & \multicolumn{2}{l}{Cramer-von Mises} \\
		\cmidrule{3-8}    &      & Statistics & P-value & Statistics & P-value & Statistics & P-value \\
		\hline
ComGBII & 0.998 & 0.014 & 1.000 & 0.639 & 1.000 & 0.072 & 1.000 \\
GBIIG & 0.998 & 0.013 & 0.976 & 0.658 & 0.923 & 0.077 & 0.901 \\
BIIG  & 0.996 & 0.022 & 0.829 & 1.811 & 0.783 & 0.173 & 0.825 \\
BG    & 0.998 & 0.015 & 0.956 & 0.729 & 0.921 & 0.081 & 0.936 \\
IBG   & 0.998 & 0.015 & 0.907 & 0.638 & 0.933 & 0.078 & 0.904 \\
PG    & 0.998 & 0.016 & 0.914 & 0.850 & 0.842 & 0.098 & 0.862 \\
IPG   & 0.995 & 0.023 & 0.863 & 2.302 & 0.748 & 0.244 & 0.780 \\
		\hline
		    WIW   &  -     & 0.021 & 0.222 & 1.159 & 0.284 & -  &- \\
		PIW   &    -   & 0.021 & 0.226 & 1.156 & 0.285 & -  &- \\
		IBW   &   -    & 0.021 & 0.216 & 1.160  & 0.283 & - & - \\
		WIP   &   -    & 0.021 & 0.210  & 1.318 & 0.227 &-    &- \\
		IBIP   &   -    & 0.021 & 0.211 & 1.327 & 0.224 & -  & -\\
		IBB   &  -     & 0.015 & 0.636 & 0.711 & 0.550  & -  & - \\
		\bottomrule
	\end{tabular*}%
	\label{tab:gof - danish}%
	\begin{tablenotes}
		\item \small The bootstrap p-values are computed using parametric bootstrap with 2000 simulation runs.
		\item \small The results of competing models are obtained in \cite{grun2019extending}. The correlation R of QQ-plots and Cramer-von Mises are not reported in their study.
	\end{tablenotes}
\end{table}%

\begin{table}[tp!]
	\centering
	\caption{{Danish fire insurance claims: summary of the results for VaR and TVaR at the 0.95 and 0.99 confidence level, obtained for the composite GBII models and previously fitted composited models.}}
%
	\begin{tabular*}{\hsize}{@{}@{\extracolsep{\fill}}lcccccccc@{}}
		\toprule
		\multirow{2}[1]{*}{Model}  & \multicolumn{4}{c}{VaR}  & \multicolumn{4}{c}{TVaR} \\
		\cmidrule{2-9}         &    0.95   & Diff.\%         &    0.99   & Diff..\%          &    0.95   & Diff..\%         &    0.99   & Diff.\%\\
		\hline
	Empirical & 8.41  &- & 24.61 &  -     & 22.16 &     -  & 54.60 & - \\
	\hline
ComGBII & 8.20  & -2.42 & 25.19 & 2.36  & 27.09 & 22.28 & 83.21 & 52.39 \\
GBIIG & 8.28  & \textbf{-1.54} & 25.73 & 4.54  & 28.03 & 26.52 & 87.14 & 59.58 \\
BIIG  & 8.16  & -2.98 & 24.97 & 1.47  & 26.77 & 20.84 & 81.98 & 50.14 \\
BG    & 8.27  & -1.59 & 25.67 & 4.31  & 27.92 & 26.02 & 86.65 & 58.69 \\
IBG   & 8.24  & -1.96 & 25.34 & 2.93  & 27.27 & 23.08 & 83.83 & 53.52 \\
PG    & 8.16  & -2.87 & 24.96 & 1.43  & 26.72 & 20.60 & 81.69 & 49.61 \\
IPG   & 8.08  & -3.90 & 24.49 &\textbf{ -0.49} & 25.99 & 17.31 & 78.80 & 44.32 \\
		\hline
	WIW   & 8.02  & -4.60 & 22.77 & -7.49 & 22.64 & 2.19  & 63.86 & 16.95 \\
	PIW   & 8.02  & -4.60 & 22.79 & -7.41 & 22.67 & 2.32  & 64.00 & 17.21 \\
	IBW   & 8.01  & -4.71 & 22.73 & -7.65 & 22.59 & 1.96  & 63.67 & 16.60 \\
	WIP   & 8.03  & -4.48 & 22.64 & -8.02 & 22.38 & \textbf{1.02}  & 62.65 & \textbf{14.74} \\
	IBP   & 8.03  & -4.48 & 22.65 & -7.98 & 22.39 & 1.06  & 62.69 & 14.81 \\
	IBB   & 8.22  & -2.22 & 25.13 & 2.10  & 26.88 & 21.33 & 82.15 & 50.45 \\
		\bottomrule
	\end{tabular*}%
	\label{tab:TVaR-competing models}%
\end{table}%

\begin{figure}[tp!]
	\centering
	\includegraphics[scale=0.6]{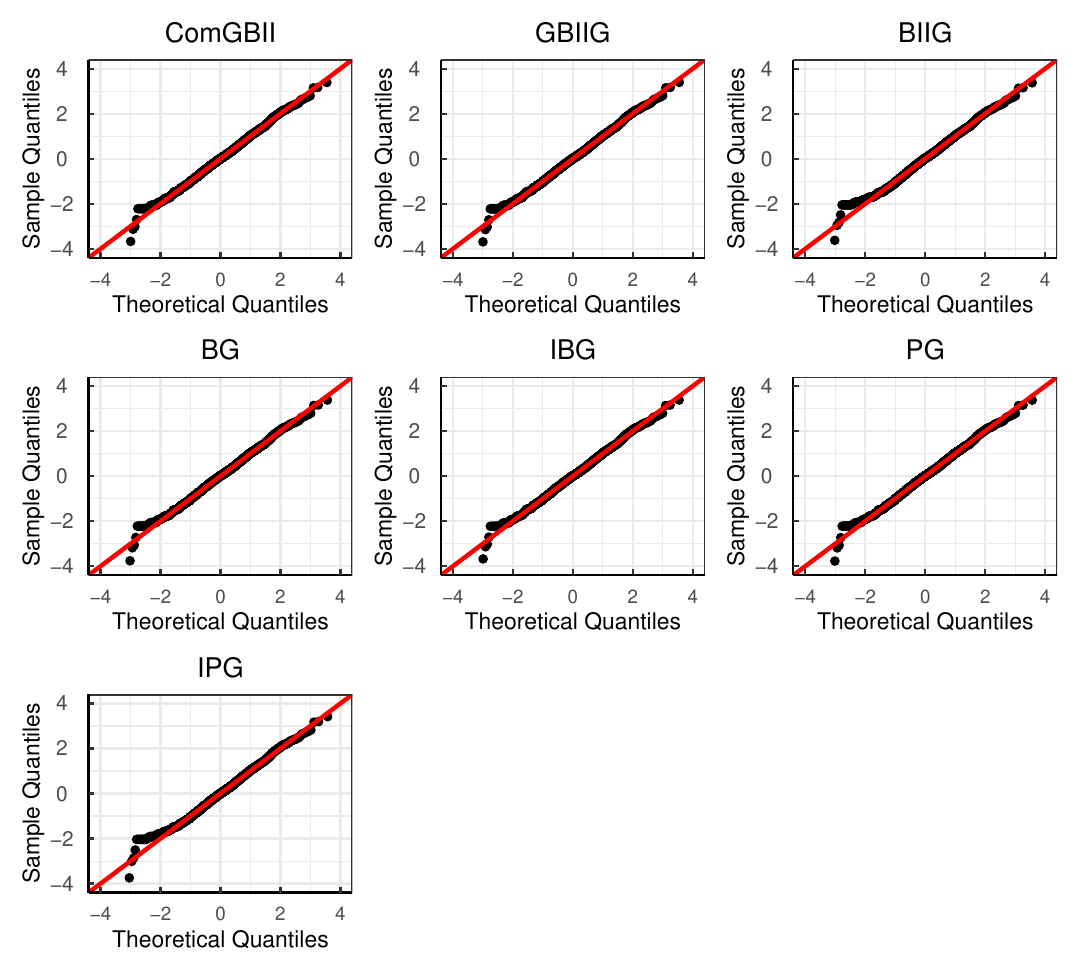}	
	\caption{Danish fire insurance claims: QQ-plots of the empirical quantiles against the estimated model quantiles.}
	\label{fig:QQ-plot-danish}
\end{figure}

In Figure \ref{fig:QQ-plot-danish}, the QQ-plots of the empirical quantiles against
the estimated quantiles of quantile residuals from the seven proposed models are
given. The correlation coefficients R of these QQ-plots are also given in
Table \ref{tab:gof - danish}: R measures the degree of linearity in the QQ-plot and hence also the
goodness-of-fit with respect to the corresponding model.
One can see that the plots
in Figure \ref{fig:QQ-plot-danish} also indicate that the proposed seven distributions all give good
fits in the sense that the points corresponding
to the theoretical and empirical quantiles do not deviate much
from the $45^{\circ}$ straight line.

Finally, Table \ref{tab:TVaR-competing models} reports the empirical and estimated values of
the VaR and TVaR at confidence levels of 95\% and 99\%.
The percentage of
variation of each estimated VaR and TVaR, with respect to the empirical
VaR and TVaR, is also reported. 
{
	When examining the VaR estimates at a 95\% confidence level, the GBIIG  model performs the best, with a 1.54\% underestimation of the empirical VaR. Similarly, at a 99\% confidence level, the IPG model is the most accurate, with a slight underestimation of 0.49\%. 
	However, in the case of TVaR estimates, our proposed models do not outperform the competing models. The WIP model consistently provides the best performance across different confidence levels. This can be attributed to the thicker tail features of the proposed models, which lead to higher prediction results compared to the competing composite distribution.
}


\clearpage

\subsection{Medical insurance data set}\label{section:application3}
\begin{figure}[htb!]
	\centering
	\includegraphics[scale = 0.7]{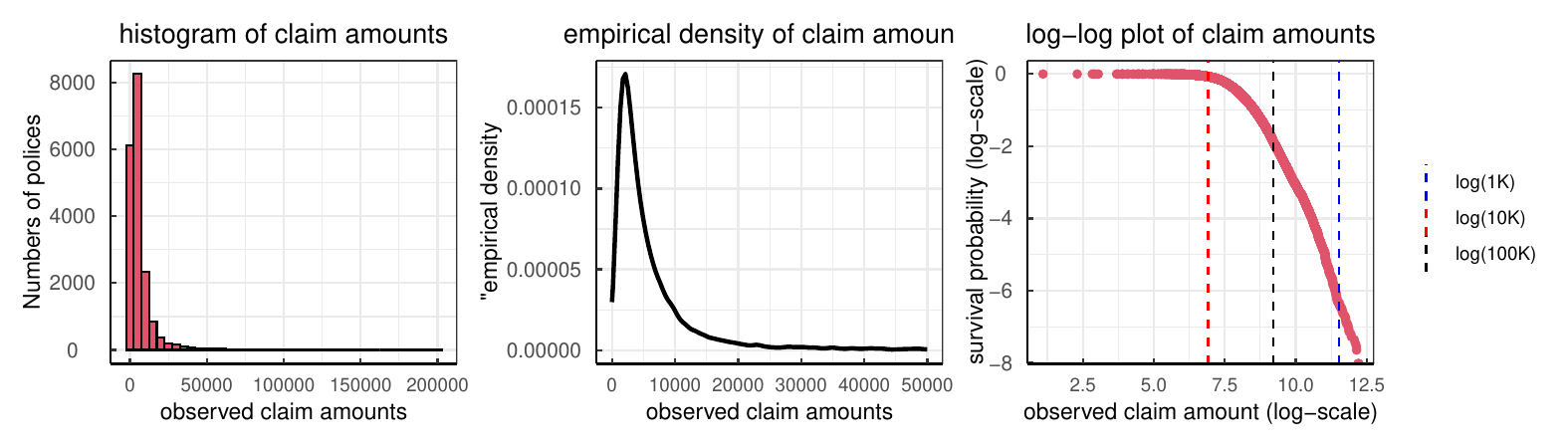}
	\caption{\textit{Left panel}: histogram of the (positive) claim amounts of individual polices;
		\textit{Middle panel}: empirical density (upper-truncated at 50,000); 
		\textit{Right panel}: log-log plot of observed claim amounts.}
	\label{fig-medical-hist}
\end{figure}

\begin{figure}[htb]
	\centering
	\includegraphics[scale = 0.55]{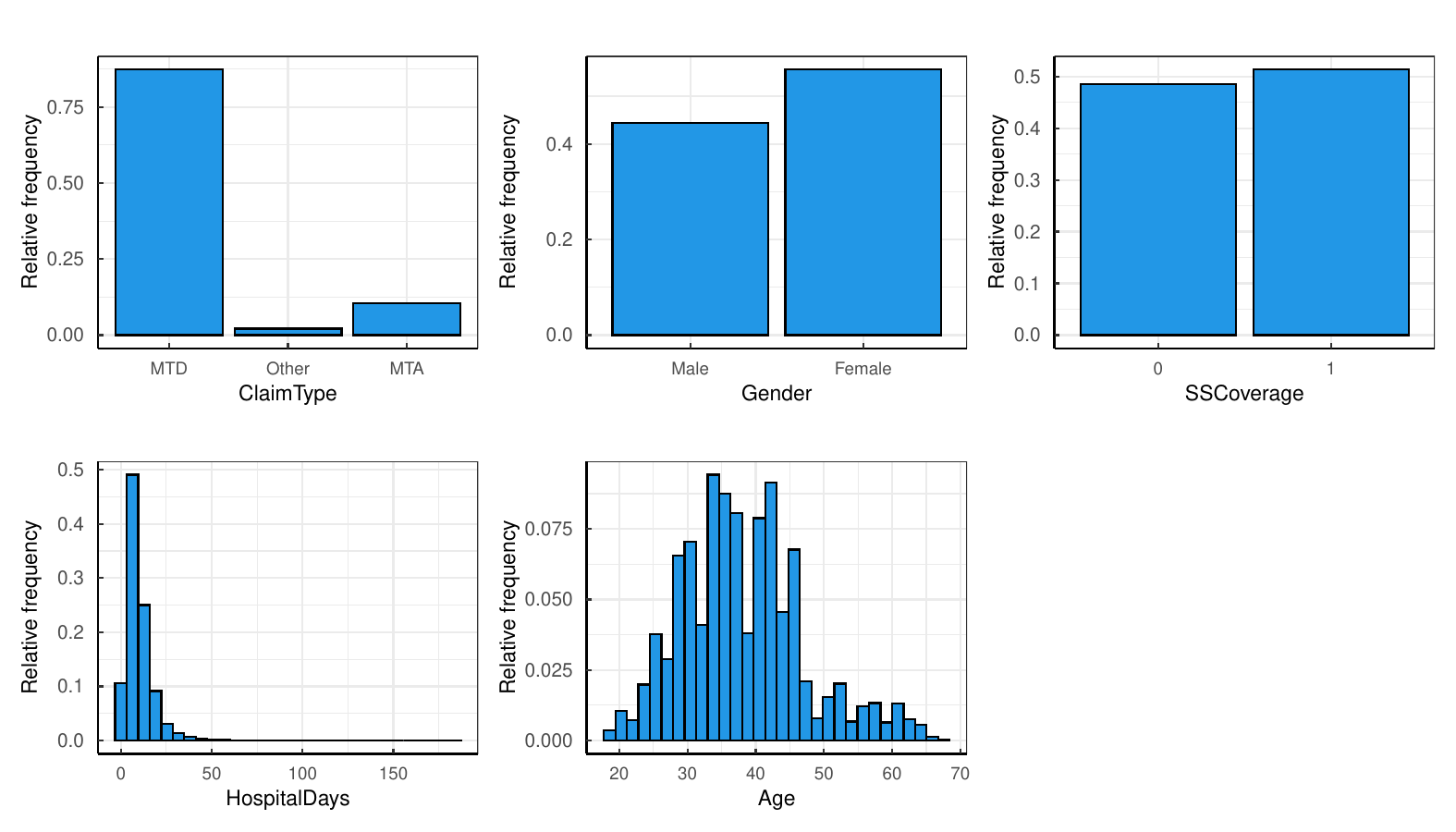}
	\caption{Relative frequency of the covariates \textit{Gender}, \textit{Age}, \textit{SSCoverage}, \textit{HospitalDays} and \textit{ClaimType}.}
	\label{fig-medical-frequency}
\end{figure}

As the second example, we consider a medical insurance data set that was kindly provided by a major insurance company operating in China and concerns a short-term medical insurance that covers individual inpatient claim data.
The data contains 19,110 policies between 2014 and 2016.
Each claim records
the positive inpatient claim amount, 
and several covariates, including \textit{Gender}, \textit{Age}, \textit{SSCoverage}, \textit{HospitalDays} and \textit{ClaimType}.
The summary statistics of these variables are shown in Table \ref{tab-summary-2} and 
the histogram, the empirical density and the log-log plot of claim amounts are given in Figure \ref{fig-medical-hist}.
The empirical density is unimodal and highly skewed.
The log-log plot seems asymptotically linear for the largest 10\% of
the claim amounts which indicates the tail is heavy-tailed.
Figure \ref{fig-medical-frequency} shows how the covariates from Table \ref{tab-summary-2} are distributed in the medical insurance data-set.
More than half of the policyholders (51.41\%) have the social medical insurance or public medical insurance coverage, 
and more than half of the policyholders (55.63\%) are female.
Most policyholders (87.45\%) have applied for insurance claims for MTD, 10.45\% for MTA and only 
2.10\% for other level which includes the disability from disease (DSD), death from disease (DED), disability from accident (DSA), death from accident (DEA) and major disease (MD).
Almost
all policyholders (93.73\%) are aged between 25 and 60, which means that there are few young and
old inpatients in the insurance portfolio.
Most of the policyholders in the insurance portfolio have been hospitalized less than 30 days (97.93\%). 

\begin{figure}[htb!]
	\centering
	\includegraphics[scale=0.5]{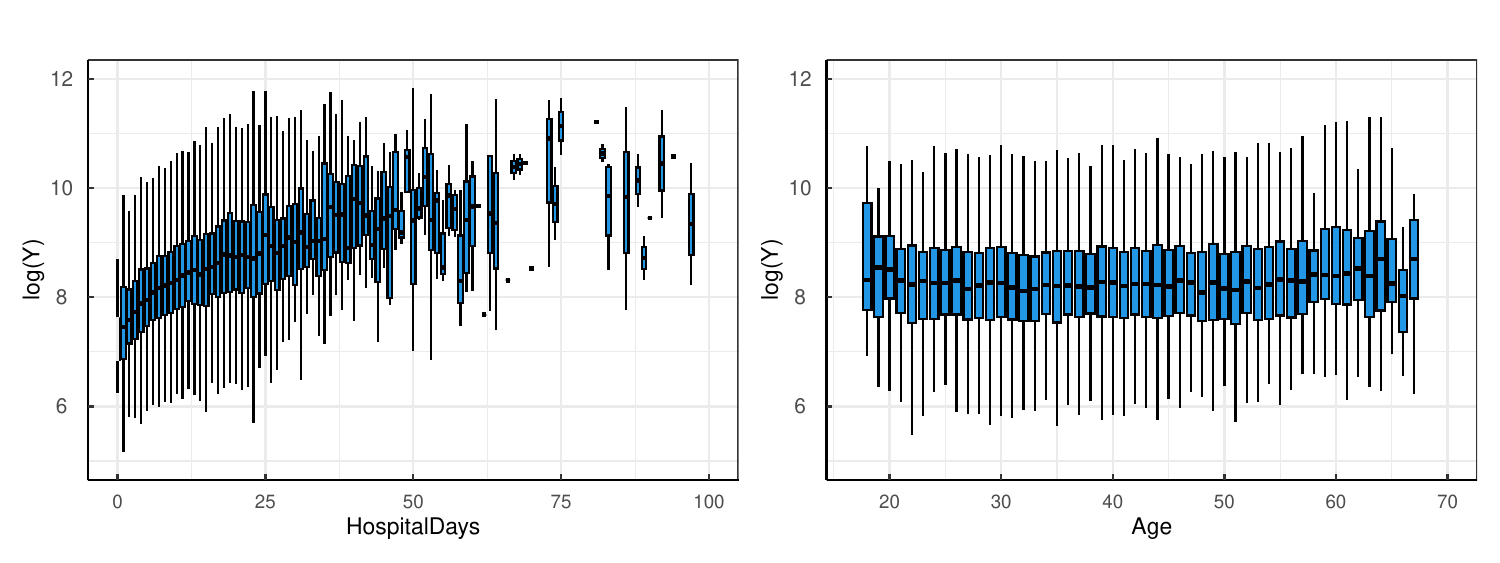}
	\caption{{Boxplot of $\log(Y)$ with respect to the two continuous covariates: \textit{HospitalDays} and \textit{Age}. }}
	\label{fig-medical-nonlinear}
\end{figure}

\begin{figure}[tp]
	\centering
	\includegraphics[scale=0.5]{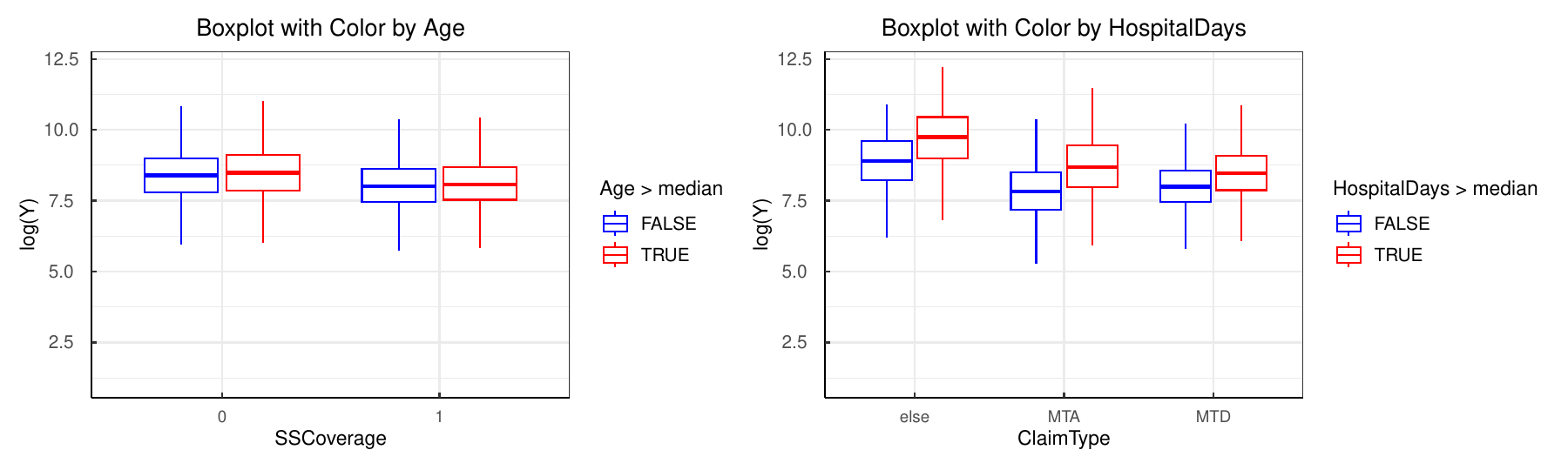}
	\caption{{Boxplot of $\log(Y)$ with respect to the interaction effect of two covariates. The x-axis represents the levels of one variable, while the color of each box corresponds to the levels of the other variable. }
	}
	\label{fig-medical-interaction}
\end{figure}

\begin{figure}[htb!]
	\centering
	\includegraphics[scale=0.6]{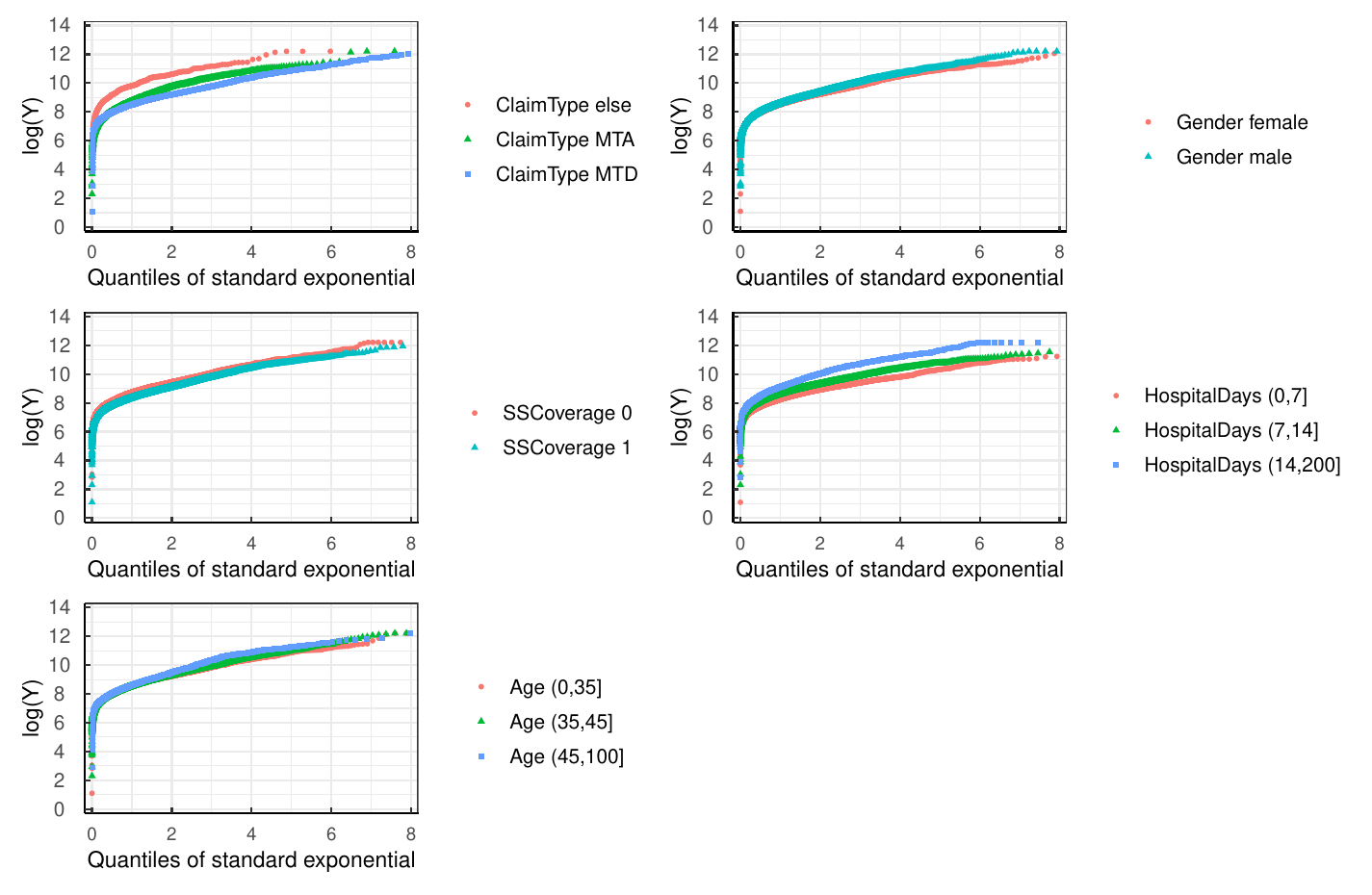}
	\caption{{Pareto QQ-plots of claim amount for different intervals of levels for the covarites.}}
	\label{fig-pareto-qq-covarites}
\end{figure}

{
To investigate the nonlinear effects of two continuous covariates on the claim amounts, Figure \ref{fig-medical-nonlinear} shows a boxplot of  log-transformed claim amount with respect to \textit{Age} and \textit{HospitalDays}.
One can see that the linear effects of these two covariates are sufficient to characterize the log-transformed claim data.
To further explore the interaction effects of these covariates, 
Figure \ref{fig-medical-interaction} shows a boxplot of log-transformed claim amount with respect to \textit{Age} and \textit{SScoverage} in the left penal, and with respect to \textit{HospitalDays} and \textit{ClaimType} in the right penal.
The effect of age on insurance claims varies depending on whether an individual has paid social insurance or not.
It also suggests a significant interaction effect, where the relationship between $\log(Y)$ and \textit{HospitalDays} changes across the levels of \textit{ClaimType}. 
These preliminary findings 
highlight the importance of considering the interaction effect between these covariates in the regression models.	
Moreover, 
Figure \ref{fig-pareto-qq-covarites} presents the Pareto QQ-plots of the individual claim amount for various levels of five covariates. These plots reveal an increase in the extreme value index as the \textit{Age} and \textit{HospitalDays} variables increase, as the slopes of these plots at the largest claim levels provide a graphical inspection of the extreme value index, see \cite{beirlant2004}. 
Also, the plots demonstrate that the tail behaviors of the insurance claims differ across the levels of categorical variables, such as \textit{ClaimType}, \textit{Gender}, and \textit{SSCoverage}. 
We can see that these effects of the covariates on the tail behavior have the same direction as the estimated regression coefficients reported in the GLM Gamma, see Table \ref{tab-coef-medical}, which indicates that the effects of the covariates on the body and the tail are almost in the same direction.
These empirical findings motivate us to explore the proposed composite regression models, as it is consistent with the assumption of the proposed model that 
the covariates affect (exponential linearly) the claims in the head part, the claims in the tail part, and the composite threshold simultaneously (in the same direction), see \eqref{eq: modelling setting}-\eqref{eq: varying-threshold}.
}


\begin{table}[htb!]
	\small
	\centering
	\caption{Medical insurance claim data set: description of variables.}
	\renewcommand\arraystretch{1.5}
	\begin{tabular*}{\hsize}{@{}@{\extracolsep{\fill}}lll@{}}
		\toprule
		\textbf{Variables} & \textbf{Type} & \textbf{Description}\\
		\hline
		Claim amount&	Continuous	&inpatient (positive) claim amount of a patient \yen 3-200,000 \\  
		Age&	Continuous&	inpatient's age: 18-67\\
		Gender &	Categorical & male, female\\
		SSCoverage &	Categorical & social medical insurance or public medical insurance coverage: 0-1\\
		HospitalDays &	Continuous & the duration of hospitalization of a inpatient: 0-184 days\\
		ClaimType &	Categorical & \tabincell{l}{reasons for claiming the medical insurance claim:\\ medical treatment from disease (MTD), \\ medical treatment from accident (MTA), other}\\
		\bottomrule
	\end{tabular*}
	\label{tab-summary-2}
\end{table}

For our analysis, we split the data set $\mathcal{L}(Y_{i},\bm{x}_i)_{0\le i\le n}$ into a training set $\mathcal{U}$ that is used
for model fitting, and a testing set $\mathcal{V}$ which we (only) use for an out-of-sample
analysis in proportion 60:40.  
We fit the distribution of claim amounts by incorporating covariates through the ComGBII regression and its sub-models discussed above (GBIIG, BIIG, BG, IBG, PG, IPG),
illustrating how the covariates and the interactions have significant effect on the claim amount paid for each insurance policy.
{The results are also compared with the GBII regression model, the generalized log-Moyal regression model (GlogM) discussed in \cite{bhati2018generalized},
the Burr regression discussed in \cite{beirlant1998burr},  
FMR (GA),
LRMoE (GA),
composite Gamma-Pareto regression,
TPM,
as well as the log-link GLMs and generalized additive models (GAMs) with Gamma distributional assumption.
Specifically, the covariates are included in the GBII model to capture their effect on the scale parameter $\mu$, where a log link is used. 
In the GAMs, the covariates \textit{HospitalDays} and \textit{Age} are modeled using cubic splines. 
Additionally, the interaction effect between \textit{HospitalDays} and \textit{ClaimType}, as well as \textit{Age} and \textit{SSCoverage}, are included in all competing models to account for their potential impact on the claim amounts.}

\begin{figure}[hbt!]
	\centering
	\includegraphics[scale = 0.5]{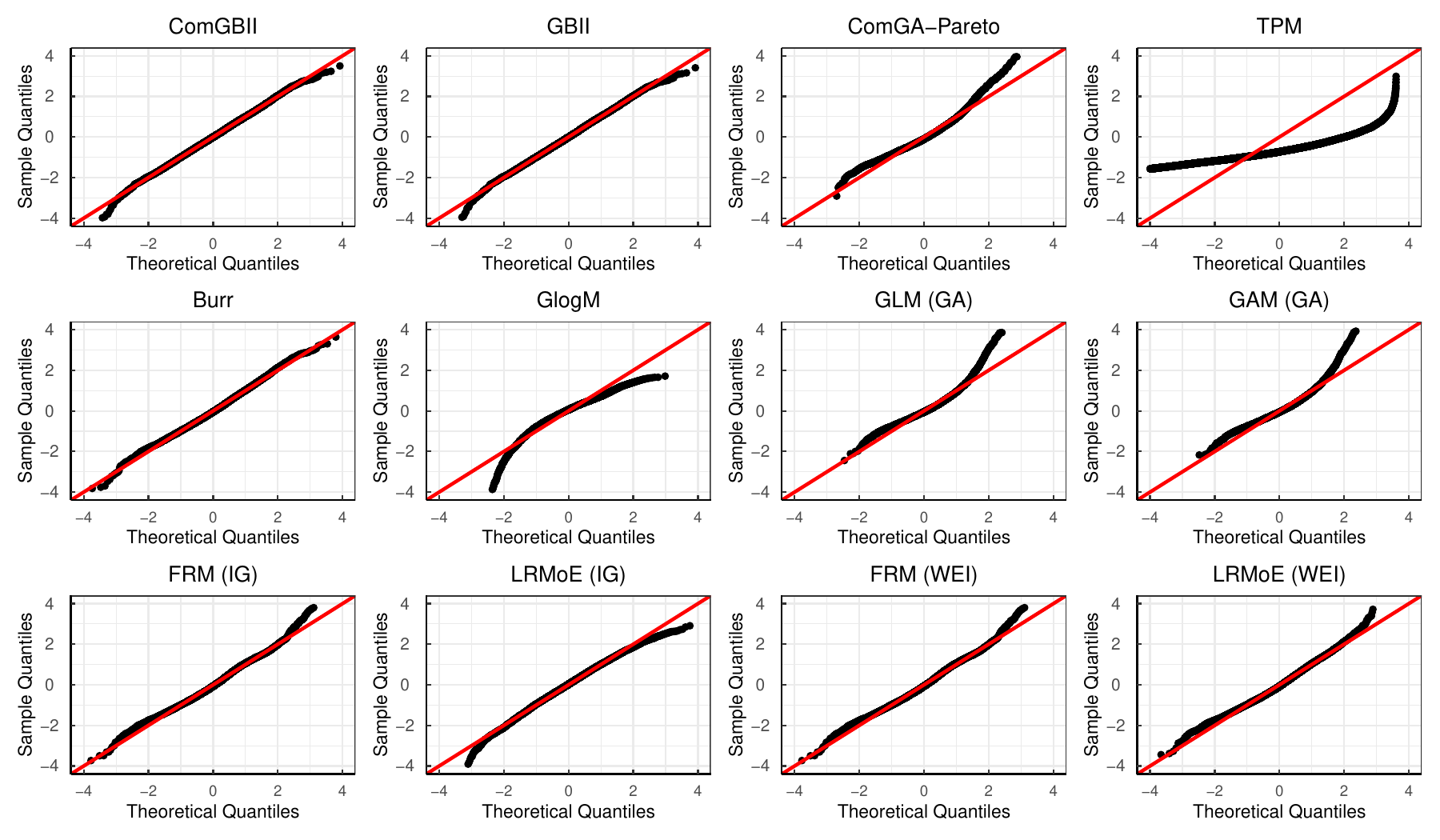}
	\caption{\red{Training data set: normal QQ-plots of quantile residuals $r_i$ from the competing models.} 
	}
	\label{fig-medical-qq}
\end{figure}

The goodness-of-fit measures, including NLL, AIC, and BIC values, for various models fitted to the claim amounts data, are reported in Table \ref{tab:goodness - medical}. The ComGBII model shows the best fit in terms of AIC, followed by the GBII, IBG, and GBIIG models, while GBII model shows the best fit items of BIC, followed by the PG and IBG models. 
Other competing models, such as GLMs, GAMs, FRM, LMoE, TPM, among others, exhibit much higher BIC values than the ComGBII and its sub-models (GBIIG, BIIG, BG, IBG, PG, IPG), suggesting a poorer fit. 
Although the GAMs perform better than the GLMs in terms of AIC and BIC, which implies that the non-linear effect of the continuous covariates on the claim amounts should be taken into account for improving the goodness of fit, the use of spline functions reduces the interpretability of the model. 
Further investigation into the non-linear effects of the covariates on the response variable in the proposed composite regression model is necessary for future research.
To demonstrate the goodness of fit of the proposed models, we also provide in Figure \ref{fig-medical-qq} the QQ-plots of the quantile residuals for several selected competing models.
It clearly indicates a preference for the ComGBII and GBII regression models, which provide a considerably good fit for estimating the entire distribution of claim amounts in general. The upper tails of the distribution can be better fitted by applying the ComGBII model than the GBII model. 
\red{
Despite the use of heavy-tailed distributions, specifically the Weibull and Inverse Gaussian (IG), as the mixture components, the FRM and LMoE models exhibit limited efficacy in accurately capturing claim data across a broad range of claim amounts, especially in comparison to the Burr, GBII and ComGBII models.
}

\begin{table}[hbt!]
	\centering
	\caption{\red{Medical insurance claim data set (in-sample): model selection.}}
	\begin{tabular*}{\hsize}{@{}@{\extracolsep{\fill}}lcccccc@{}}
			\toprule
			\textbf{Models} &      \textbf{Npars} &        \textbf{NLL} &        \textbf{AIC} &    \textbf{Ranking} &        \textbf{BIC} &    \textbf{Ranking} \\
			\hline
ComGBII & 16    & \textbf{107586.30} & \textbf{215204.50} & 1     & 215321.80 & 4 \\
GBIIG & 15    & 107594.90 & 215219.80 & 4     & 215329.70 & 6 \\
BIIG  & 14    & 107607.70 & 215243.40 & 8     & 215346.00 & 8 \\
BG    & 14    & 107598.00 & 215224.00 & 6     & 215326.60 & 5 \\
IBG   & 14    & 107594.80 & 215217.70 & 3     & 215320.30 & 3 \\
PG    & 13    & 107598.70 & 215223.40 & 5     & 215318.70 & 2 \\
IPG   & 13    & 107605.80 & 215237.60 & 7     & 215332.80 & 7 \\
\hline
GBII  & 13    & 107592.40 & 215210.80 & 2     & \textbf{215306.10} & 1 \\
GLM (GA) & 11    & 110125.00 & 220271.90 & 17    & 220352.50 & 16 \\
GAM (GA) & 32    & 108418.30 & 216907.80 & 14    & 217168.90 & 13 \\
FRM (IG) & 23    & 107766.20 & 215578.40 & 11    & 215746.90 & 10 \\
FRM (WEI) & 23    & 107738.90 & 215523.80 & 10    & 215692.40 & 9 \\
LMoE (IG) & 14    & 108680.60 & 217389.20 & 15    & 217491.80 & 14 \\
LMoE (WEI) & 14    & 108170.10 & 216368.30 & 12    & 216470.90 & 11 \\
ComGA-Pareto & 25    & 109748.40 & 219546.80 & 16    & 219730.00 & 15 \\
TPM   & 25    & 108425.00 & 216899.90 & 13    & 217083.20 & 12 \\
Burr  & 12    & 107642.80 & 215309.60 & 9     & 237809.60 & 17 \\
GlogM & 12    & 110652.90 & 221329.90 & 18    & 243829.90 & 18 \\
			\bottomrule
		\end{tabular*}  
	\label{tab:goodness - medical}%
	\begin{tablenotes}
		\item \small 	Best performance is in boldface. The Npars of GAMs is determined by the model's degrees of freedom.
	\end{tablenotes}
\end{table}%

{The results of the fitted ComGBII model with all covariates, along with GBII regression and the GLMs, are presented in Table \ref{tab-coef-medical}. Standard errors were obtained from a normal approximation using the implied Fisher matrix based on the observed Hessian matrix from the numerical optimization, as described in Section \ref{sec: regression}. Notably, most of the covariate levels in the ComGBII model exhibit high statistical significance (p-values less than 0.001), indicating their importance in explaining both the body and tail of claim distribution. The exception is the \textit{Gender} variable, which shows significant effect at the 0.05 level of significance for the ComGBII model (p-value is 0.04). 
Interestingly, the signs of the estimated regression coefficients of most variables across all three models are consistent, implying that the interpretability of the regression coefficients from the ComGBII model is comparable to that of the GLMs, although the level of statistical significance differs. 
For example, the variable \textit{SSCoverage} shows significant negative effects in the ComGBII, while it is not significant in the GLMs.
Furthermore, we observe that the estimated $\hat{p}_1\hat{\tau}_1 < 1$ and $\hat{p}_2\hat{\tau}_2 > 1$ in the ComGBII model, indicating the (theoretical) mean exits for the body distribution, while it does not exist for the tail distribution. 
The VaR measures of the ComGBII model can be used for prediction, which are proportional to some exponential transformation of the linear combinations of covariates. This provides an intuitive interpretation for insurance classification ratemaking and reserving when modelling more extreme claim data. For example, our estimation results from the ComGBII model suggest that the estimated VaR of inpatients in the male group is $\exp(0.03) = 1.03$ times higher than that of the female group.
}

\begin{table}[hbt!]
	\centering
	\caption{{Medical insurance claim data set: estimation results of ComGBII, GLMs and GBII models.}}
	\begin{tabular*}{\hsize}{@{}@{\extracolsep{\fill}}ccccccc@{}}
		\toprule
		\multirow{2}[0]{*}{Parameters} & \multicolumn{2}{c}{ComGBII} & \multicolumn{2}{c}{GBII} & \multicolumn{2}{c}{GLMs (GA)} \\
		\cline{2-7}
		& Esimates & S.E.  & Esimates & S.E.  & Esimates & S.E. \\
		\hline
(Intercept) & 8.91*** & 0.11  & 8.99*** & 0.10  & 9.38*** & 0.13 \\
ClaimType\_MTA & -1.33*** & 0.09  & -1.35*** & 0.09  & -1.19*** & 0.12 \\
ClaimType\_MTD & -1.30*** & 0.08  & -1.31*** & 0.08  & -1.42*** & 0.11 \\
Gender\_Male & 0.03  & 0.02  & 0.03 & 0.02  & 0.03  & 0.02 \\
SSCoverage\_1 & -0.22** & 0.07  & -0.22** & 0.07  & -0.15 & 0.10 \\
HospitalDays & 0.02*** & 0.00  & 0.02*** & 0.00  & 0.02*** & 0.00 \\
Age & 0.01*** & 0.00  & 0.01*** & 0.00  & 0.01*** & 0.00 \\
SSCoverage\_1:Age & -0.003*    & 0.00  & -0.004*    & 0.00  & -0.005     & 0.00 \\
ClaimType\_MTA:HospitalDays & 0.02*** & 0.00  & 0.02*** & 0.00  & 0.02*** & 0.01 \\
ClaimType\_MTD:HospitalDays & 0.02*** & 0.00  & 0.02*** & 0.00  & 0.03*** & 0.00 \\
		\hline
    $p_1$    & 4.21  & 0.48  & -     & -     & -     & - \\
$p_2$    & 1.32  & 0.12  & -     & -     & -     & - \\
$\tau_1$  & 0.05  & 0.01  & -     & -     & -     & - \\
$\tau_2 $ & 1.78  & 0.17  & -     & -     & -     & - \\
$\nu_1$   & 0.50  & 0.08  & -     & -     & -     & - \\
$\nu_2$   & 2.04  & 0.18  & -     & -     & -     & - \\
$p $    & -     & -     & 1.64  & 0.33  & -     & - \\
$\tau$   & -     & -     & 1.31  & 0.37  & -     & - \\
$\nu$    & -     & -     & 1.44  & 0.42  & -     & - \\
$\phi$   & -     & -     & -     & -     & 1.51  & - \\
		\bottomrule
	\end{tabular*}  
	\begin{tablenotes}
					\item \small  *, ** and *** represents the p-value $< 0.01$, $< 0.001$ and $0$.
					\item $p_1,p_2, \tau_1, \tau_2, \nu_1, \nu_2$ represent the shape parameters in ComGBII. $p,\tau,\nu$ represent the shape parameters in GBII models. $\phi$ denotes the dispersion parameter in GLMs (GA).
	\end{tablenotes}
	\label{tab-coef-medical}
\end{table}

\begin{figure}[htbp]
	\centering
	\begin{minipage}{1\linewidth}
		\centering
		\includegraphics[width=1\linewidth]{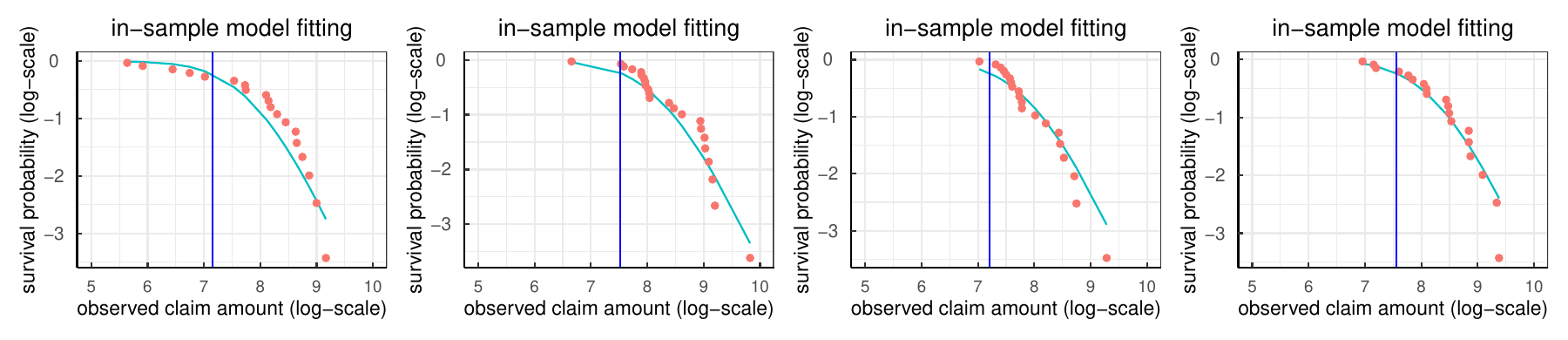}
	\end{minipage}
	\begin{minipage}{1\linewidth}
		\centering
		\includegraphics[width=1\linewidth]{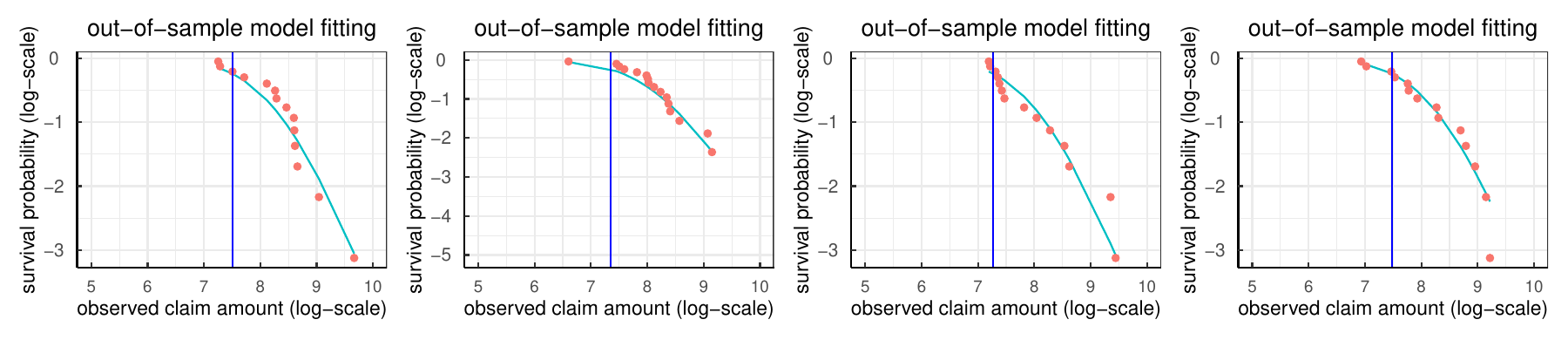}
	\end{minipage}
	\caption{{empirical (red dots) vs. fitted (green dots) log-log plots of claim amounts for randomly selected four risk classes on the training data (\textit{Top}) and testing data (\textit{Bottom}). The estimated threshold $\hat{u}(\bm{x}_i)$ based on the risk features in the fitted ComGBII distribution is marked as a blue vertical line.} }
\label{fig-splicing-class}
\end{figure}

In order to investigating the in-sample and out-of-sample performance of the proposed ComGBII regression,
Figure \ref{fig-splicing-class} shows empirical (red dots) vs. fitted (green dots) log-log plots of claim amounts for randomly selected four risk classes on the training data and testing data respectively.
The estimated threshold $\hat{u}(\bm{x}_i)$ in the fitted ComGBII distribution is marked as a blue vertical line. 
{The threshold values across all individuals range from 1,040 (minimum) to 2,146,521 (maximum).
The empirical distributions in different risk classes are highly consistent with the fitted distributions above the threshold, which suggests that the ComGBII model performs well in both in-sample and out-of-sample scenarios.}


\begin{figure}[htb]
	\centering
	\includegraphics[scale = 0.5]{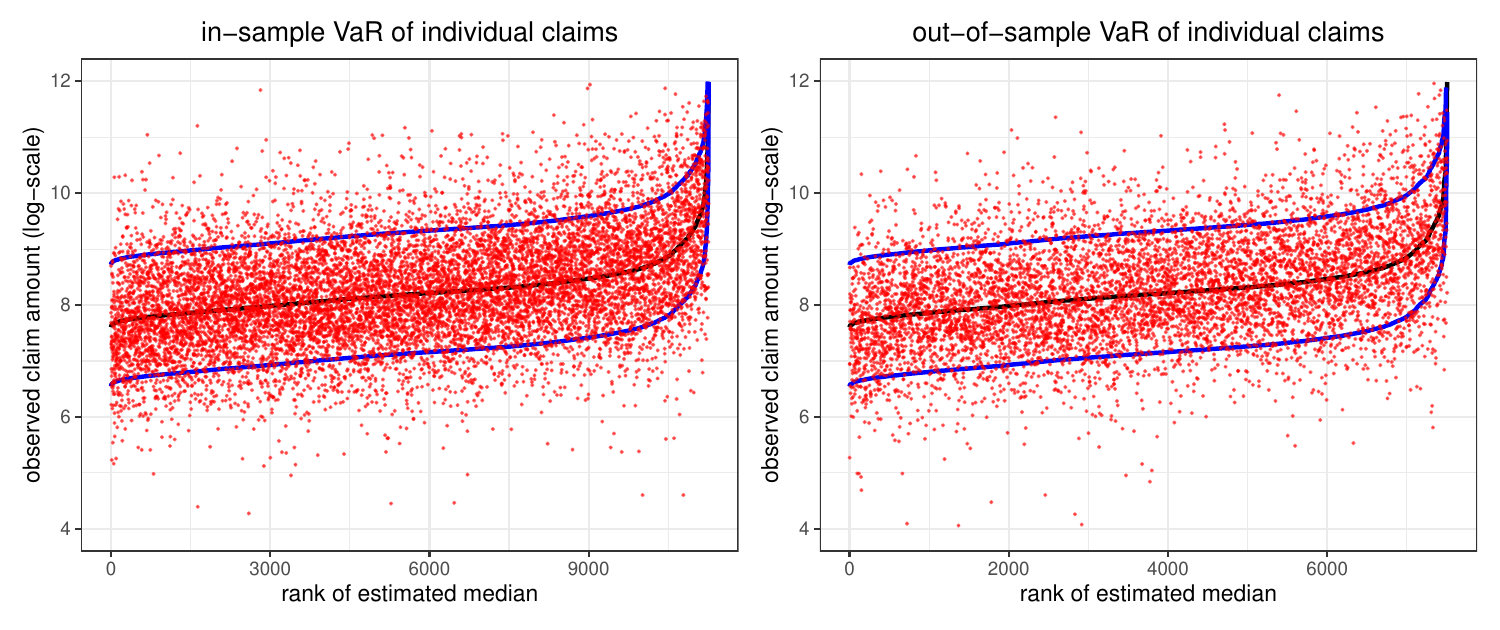}
	\caption{
		Estimated $\text{VaR}_q(Y_i;\bm{x}_i)$ of ComGBII regression model at quantile level 
		$q:10\%; 50\%; 90\%$ (blue, black, blue) on the training data $\mathcal{U}$ (\textit{Left panel}) and testing data $\mathcal{V}$ (\textit{Right panel}).   The red dots show the corresponding sample observations (realizations) $Y_i$; the x-axis orders the claims w.r.t. the estimated median $\text{VaR}_{50\%}(Y_i;\bm{x}_i)$ (in black).}
	\label{fig-medical-VaR}
\end{figure}

Table \ref{tab-medical-predictive} compares the in-sample and out-of-sample forecast performance of the different models investigated on 
the predicted $\hat{y}_{i,q}(\bm{x}_i)=\text{VaR}_q(y_i;\bm{x}_i)$ for individual risk features $\bm{x}_i$ over the six quantile level $q$ with the rang $\left[0.2,0.8\right]$\footnote{
We use the estimated VaR measures over different quantile levels for prediction instead of estimated mean as the theoretical mean does not exist for all individuals in this data set.
}.
We evaluate the mean squared error (MSE) for the competing models, on the training data and testing data respectively, which is give by:
\begin{equation}
	\label{eq-mse1}
	\text{MSE}_q = \sum_{(y_{i},\bm{x}_i)\in \left\{\mathcal{U},\mathcal{V}\right\}}\left[y_i - \hat{y}_{i,q}(\bm{x}_i)\right]^2,
\end{equation}
{This metric is used to measure the estimation accuracy in \cite{baione2019individual} by calculating the differences between the values predicted based on the estimator of the quantile with different probability levels and the values observed.}
We note that ComGBII and BIIG regression models show a better in-sample and out-of-sample model performance than other models.
Figure \ref{fig-medical-VaR} shows the estimated $\text{VaR}_q(Y_i;\bm{x}_i)$ at quantile level 
$q:10\%; 50\%; 90\%$ (blue, black, blue) on the training data $\mathcal{U}$ and testing data $\mathcal{V}$ for the ComGBII model. 
The individual sample observations are ordered w.r.t. the estimated median in black. 
It can be observed that this order monotonicity for all quantiles.
The red dots show the corresponding observed claim amount, most of which fall in the interval between the estimated $\text{VaR}_q(Y_i;\bm{x}_i)$  at $q=10\%$ and $q=90\%$.

\begin{table}[hbt!]
	\centering
	\caption{{In-sample and out-of-sample $\text{MSE}_{q}, q\in \left[0.2,0.8\right]$, on the training data $\mathcal{U}$ and testing data $\mathcal{V}$ of
		nine regression models.}}
	\setlength{\tabcolsep}{2mm}{
\begin{tabular}{lccccc|ccccc}
	\toprule
\multirow{2}{*}{	\textbf{Models}/ $q$} 
	&                         \multicolumn{ 5}{c|}{\textbf{in-sample} $\text{MSE}_{q}$}&                             \multicolumn{ 5}{c}{\textbf{out-of-sample}  $\text{MSE}_{q}$} \\
\cmidrule{2-11}    
 &       20\% &        40\%  &        50\% &        60\%  &        80\%  &         20\% &        40\%  &        50\% &        60\%  &        80\% \\
	\hline
ComGBII & 5.08  & \textbf{11.28} & \textbf{16.53} & 24.63 & 66.28 & 4.38  & \textbf{9.42} & \textbf{13.71} & \textbf{20.35} & 54.62 \\
GBIIG & 5.49  & 12.69 & 18.67 & 27.66 & 72.13 & 4.70  & 10.52 & 15.39 & 22.73 & 59.14 \\
BIIG  & \textbf{5.02} & 11.48 & 16.73 & 24.56 & \textbf{63.23} & \textbf{4.38} & 9.70  & 14.06 & 20.58 & \textbf{52.87} \\
BG    & 5.23  & 12.02 & 17.60 & 25.96 & 67.26 & 4.52  & 10.06 & 14.65 & 21.54 & 55.69 \\
BG    & 5.46  & 12.59 & 18.50 & 27.40 & 71.41 & 4.67  & 10.44 & 15.25 & 22.51 & 58.52 \\
PG    & 5.21  & 11.94 & 17.47 & 25.76 & 66.67 & 4.50  & 10.01 & 14.56 & 21.39 & 55.26 \\
IPG   & 5.09  & 11.65 & 17.00 & 24.97 & 64.30 & 4.42  & 9.83  & 14.26 & 20.88 & 53.66 \\
\hline
GBII  & 5.08  & 11.36 & 16.55 & \textbf{24.47} & 64.78 & 4.40  & 9.55  & 13.83 & 20.38 & 53.83 \\
GLMs (GA)    & 8.64  & 73.43 & 164.48 & 339.94 & 1407.49 & 5.74  & 44.86 & 100.13 & 206.85 & 857.39 \\
	\bottomrule
\end{tabular}  
	}
\begin{tablenotes}
			\item The results reported are scaled by $10^{-8}$.
\end{tablenotes}
\label{tab-medical-predictive}
\end{table}


{Finally, we use the Gini index to assess the model's ability to discriminate risk across samples.
The Gini index proposed by \cite{frees2011summarizing} is a more comprehensive metric for assessing the mutual advantages among various models,
and is particularly suitable for the heavy-tailed insurance data \citep{frees2014insurance, yang2018insurance}.
This index is defined as twice the area between the ordered Lorenz curve and the straight line $y=x$, while for the two sets of predictions $\hat{\bm{y}}_1(\bm{x}_i)$ and $\hat{\bm{y}}_2(\bm{x}_i)$ produced by different models, the Lorenz curve can be drawn by:
\begin{equation}
\left(\frac{\sum_{i=1}^n \hat{\bm{y}}_1(\bm{x}_i) \cdot \text{I}[R(\bm{x}_i) \leq s]}{\sum_{i=1}^n \hat{\bm{y}}_1(\bm{x}_i)}, \ \frac{\sum_{i=1}^n y_i \cdot \text{I}[R(\bm{x}_i) \leq s]}{\sum_{i=1}^n y_i} \right), s \in [0,1],
\end{equation}
where $R(\bm{x}_i) = \hat{\bm{y}}_2(\bm{x}_i) / \hat{\bm{y}}_1(\bm{x}_i)$ represents the relative predicted difference between the two models. Here, $\hat{\bm{y}}_1(\bm{x}_i)$ and $\hat{\bm{y}}_2(\bm{x}_i)$ are the estimated VaR measures at the same confidence level. Note that the results of the Gini index are consistent across different confidence levels.
The optimal model, relative to the base model, is the one with the largest Gini index among all competing models. To identify the model with the most robust performance, the maximum Gini value of each model is recorded by taking all models in turn as the base model. The "min-max" criterion is then used to determine the best model.
Based on the Gini indices reported in Table \ref{medical-gini} for the nine models, it is observed that the maximal Gini index ranges from 0 to 28.52 when using the ComGBII and its sub-models as well as GBII and GLMs (GA) as the base. The ComGBII model has the smallest maximum Gini index at 0, indicating that it is the most robust to the alternative models.
}

\begin{table}[htbp]
	\centering
	\caption{{Gini indices for the competing models.}}
	\begin{tabular}{lccccccccc|c}
		\toprule
	\multirow{2}{*}{\textbf{Base models}}&\multicolumn{9}{c|}{\textbf{Competing models}} & 	\multirow{2}{*}{\textbf{min-max}}\\
	\cline{2-10}
	& {ComGBII} & {GBIIG} & {BIIG} & {BG} & {IBG} & {PG} & {IPG} & {GBII} & {GLMs} & \\
		\hline
		ComGBII & 0.00  & -12.91 & -2.96 & -13.45 & -12.97 & -12.95 & -11.40 & -1.34 & -7.71 & \textbf{0.00} \\
		GBIIG & 13.63 & 0.00  & 8.37  & 8.22  & 10.11 & 8.68  & 8.57  & 11.07 & -8.31 & 13.63 \\
		BIIG  & 3.37  & -7.51 & 0.00  & -8.52 & -7.90 & -7.38 & -6.85 & 2.81  & -7.46 & 3.37 \\
		BG    & 13.95 & -7.77 & 8.93  & 0.00  & -8.18 & 13.28 & 9.57  & 13.22 & -7.83 & 13.95 \\
		IBG    & 13.64 & -10.04 & 8.72  & 8.59  & 0.00  & 9.11  & 8.96  & 11.32 & -8.25 & 13.64 \\
		PG    & 13.41 & -8.17 & 7.74  & -13.22 & -8.65 & 0.00  & 8.21  & 13.53 & -7.80 & 13.53 \\
		IPG   & 11.85 & -7.83 & 6.96  & -9.26 & -8.26 & -7.96 & 0.00  & 12.77 & -7.57 & 12.77 \\
		GBII   & 1.53  & -10.32 & -2.59 & -12.84 & -10.61 & -13.20 & -12.51 & 0.00  & -7.61 & 1.53 \\
		GLMs   & 28.45 & 28.52 & 28.41 & 28.43 & 28.52 & 28.44 & 28.42 & 28.41 & 0.00  & 28.52 \\
		\bottomrule
	\end{tabular}%
	\label{medical-gini}%
\end{table}%

\clearpage
\section{Conclusion}\label{section:conclusion}

In this paper, we have proposed a versatile composite distribution based on the GBII family allowing closed-form expressions
for a lot of insurance measures and have studied some of its
properties. 
This new model is design to model the entire range of loss data which is achieved by splicing two GBII distributions as the head and the tail respectively based on mode-matching method.
It also contains a wide range of insurance loss distributions provides the close-formed expressions for statistical modelling and estimation.
Finally, the proposed model is found more appropriate as compared to several other composite models as discussed in existing literature using the well-known Danish insurance loss data-set.

Another interesting contribution of this paper is that the regression modelling is discussed in 
the composite GBII distribution for presenting
high flexibility since covariates can be introduced in the scale parameter in a non-linear form, being
therefore an alternative to the heavy tailed regression model.
It not only provides a suitable fit for the entire range of data apart from the body, but capture the risk heterogeneity by allowing the varying threshold across individuals that are related to risk features.
Moreover, it can be also used as an important complement to the conventional GLMs in the non-life insurance classification ratemaking mechanism for modelling extreme losses data when the mean does not exist.

It is worth mentioning that the composite regression models we considered in this work were parametric, and it would be semiparametric or 
approach when functional forms other than the linear are
included in the scale parameters of the models.
Another extension of this work could be devoted to the regularization approach, boosting or neural network algorithm in the regression setting that allows us 
to investigate more complex effects (e.g. non-linear or interaction effect) of the covariates on the insurance losses.

\section*{Acknowledgement}
We are grateful to Prof. Liang Yang who has provided a number of comments for the improvement of this paper.
We also gratefully thank the two anonymous referees for their constructive comments and suggestions.
Zhengxiao Li acknowledges the financial support from National Natural Science Fund of China (Grant No. 72271056 and 71901064), 
 ``the Fundamental Research Funds for the Central Universities" in UIBE (Grant No. CXTD13-02),  and the University of International Business and Economics project for Outstanding Young Scholars (Grant No. 20YQ16).

\appendix
\section*{Appendices}
\addcontentsline{toc}{section}{Appendices}
\renewcommand{\thesubsection}{\Alph{subsection}}

%

\subsection{The gradients of the log-likelihood function}\label{app:Gradients}
The first order partial derivatives of the log-likelihood function $\partial \ell(\bm{\beta}, \bm{\alpha};\bm{y}) $, with respect to  $p_1, p_2, \tau_1, \tau_2, \nu_1, \nu_2$ can be calculated based on the following equations:

\begin{align}
\frac{\partial \ell(\bm{\beta}, \bm{\alpha};\bm{y})}{\partial p_1}
&=I\left[y_i\le u(\bm{x}_i;\bm{\beta})\right]\sum_{i=1}^{n}\frac{1}{r(\bm{\alpha})}\frac{\partial r(\bm{\alpha})}{\partial p_1}\nonumber\\
&+I\left[y_i\le u(\bm{x}_i;\bm{\beta})\right]\sum_{i=1}^{n}\frac{1}{f_{\text{GBII}}(y_i;\mu_{1}(\bm{x}_i;\bm{\beta}), p_1,\nu_1,\tau_1)}\frac{\partial f_{\text{GBII}}(y_i;\mu_{1}(\bm{x}_i;\bm{\beta}), p_1,\nu_1,\tau_1)}{\partial p_1}\nonumber\\
&-I\left[y_i\le u(\bm{x}_i;\bm{\beta})\right]\sum_{i=1}^{n}\frac{1}{I_{\nu_1,\tau_1}\left[\frac{p_1\nu_1-1}{p_1\nu_1+p_1\tau_1}\right]}\frac{\partial I_{\nu_1,\tau_1}\left[\frac{p_1\nu_1-1}{p_1\nu_1+p_1\tau_1}\right]}{\partial p_1}\nonumber\\
&+I\left[y_i > u(\bm{x}_i;\bm{\beta})\right]\sum_{i=1}^{n}\frac{1}{r(\bm{\alpha})-1}\frac{\partial r(\bm{\alpha})}{\partial p_1}\nonumber,
\end{align}
\begin{align}
\frac{\partial \ell(\bm{\beta}, \bm{\alpha};\bm{y})}{\partial p_2}
&=I\left[y_i\le u(\bm{x}_i;\bm{\beta})\right]\sum_{i=1}^{n}\frac{1}{r(\bm{\alpha})}\frac{\partial r(\bm{\alpha})}{\partial p_2}\nonumber\\
&+I\left[y_i > u(\bm{x}_i;\bm{\beta})\right]\sum_{i=1}^{n}\frac{1}{r(\bm{\alpha})-1}\frac{\partial r(\bm{\alpha})}{\partial p_2}\nonumber\\
&+I\left[y_i > u(\bm{x}_i;\bm{\beta})\right]\sum_{i=1}^{n}\frac{1}{f_{\text{GBII}}(y_i;\mu_{2}(\bm{x}_i;\bm{\beta}), p_2,\nu_2,\tau_2)}\frac{\partial f_{\text{GBII}}(y_i;\mu_{2}(\bm{x}_i;\bm{\beta}), p_2,\nu_2,\tau_2)}{\partial p_2}\nonumber\\
&+I\left[y_i > u(\bm{x}_i;\bm{\beta})\right]\sum_{i=1}^{n}\frac{1}{1-I_{\nu_2,\tau_2}\left[\frac{p_2\nu_2-1}{p_2\nu_2+p_2\tau_2}\right]}\frac{\partial I_{\nu_2,\tau_2}\left[\frac{p_2\nu_2-1}{p_2\nu_2+p_2\tau_2}\right]}{\partial p_2}\nonumber,
\end{align}
\begin{align}
\frac{\partial \ell(\bm{\beta}, \bm{\alpha};\bm{y})}{\partial \tau_1}
&=I\left[y_i\le u(\bm{x}_i;\bm{\beta})\right]\sum_{i=1}^{n}\frac{1}{r(\bm{\alpha})}\frac{\partial r(\bm{\alpha})}{\partial \tau_1}\nonumber\\
&+I\left[y_i\le u(\bm{x}_i;\bm{\beta})\right]\sum_{i=1}^{n}\frac{1}{f_{\text{GBII}}(y_i;\mu_{1}(\bm{x}_i;\bm{\beta}), p_1,\nu_1,\tau_1)}\frac{\partial f_{\text{GBII}}(y_i;\mu_{1}(\bm{x}_i;\bm{\beta}), p_1,\nu_1,\tau_1)}{\partial \tau_1}\nonumber\\
&-I\left[y_i\le u(\bm{x}_i;\bm{\beta})\right]\sum_{i=1}^{n}\frac{1}{I_{\nu_1,\tau_1}\left[\frac{p_1\nu_1-1}{p_1\nu_1+p_1\tau_1}\right]}\frac{\partial I_{\nu_1,\tau_1}\left[\frac{p_1\nu_1-1}{p_1\nu_1+p_1\tau_1}\right]}{\partial \tau_1}\nonumber\\
&+I\left[y_i > u(\bm{x}_i;\bm{\beta})\right]\sum_{i=1}^{n}\frac{1}{r(\bm{\alpha})-1}\frac{\partial r(\bm{\alpha})}{\partial \tau_1}\nonumber,
\end{align}
\begin{align}
\frac{\partial \ell(\bm{\beta}, \bm{\alpha};\bm{y})}{\partial \tau_2}
&=I\left[y_i\le u(\bm{x}_i;\bm{\beta})\right]\sum_{i=1}^{n}\frac{1}{r(\bm{\alpha})}\frac{\partial r(\bm{\alpha})}{\partial \tau_2}\nonumber\\
&+I\left[y_i > u(\bm{x}_i;\bm{\beta})\right]\sum_{i=1}^{n}\frac{1}{r(\bm{\alpha})-1}\frac{\partial r(\bm{\alpha})}{\partial \tau_2}\nonumber\\
&+I\left[y_i > u(\bm{x}_i;\bm{\beta})\right]\sum_{i=1}^{n}\frac{1}{f_{\text{GBII}}(y_i;\mu_{2}(\bm{x}_i;\bm{\beta}), p_2,\nu_2,\tau_2)}\frac{\partial f_{\text{GBII}}(y_i;\mu_{2}(\bm{x}_i;\bm{\beta}), p_2,\nu_2,\tau_2)}{\partial \tau_2}\nonumber\\
&+I\left[y_i > u(\bm{x}_i;\bm{\beta})\right]\sum_{i=1}^{n}\frac{1}{1-I_{\nu_2,\tau_2}\left[\frac{p_2\nu_2-1}{p_2\nu_2+p_2\tau_2}\right]}\frac{\partial I_{\nu_2,\tau_2}\left[\frac{p_2\nu_2-1}{p_2\nu_2+p_2\tau_2}\right]}{\partial \tau_2}\nonumber,
\end{align}
\begin{align}
\frac{\partial \ell(\bm{\beta}, \bm{\alpha};\bm{y})}{\partial \nu_1}
&=I\left[y_i\le u(\bm{x}_i;\bm{\beta})\right]\sum_{i=1}^{n}\frac{1}{r(\bm{\alpha})}\frac{\partial r(\bm{\alpha})}{\partial \nu_1}\nonumber\\
&+I\left[y_i\le u(\bm{x}_i;\bm{\beta})\right]\sum_{i=1}^{n}\frac{1}{f_{\text{GBII}}(y_i;\mu_{1}(\bm{x}_i;\bm{\beta}), p_1,\nu_1,\tau_1)}\frac{\partial f_{\text{GBII}}(y_i;\mu_{1}(\bm{x}_i;\bm{\beta}), p_1,\nu_1,\tau_1)}{\partial \nu_1}\nonumber\\
&-I\left[y_i\le u(\bm{x}_i;\bm{\beta})\right]\sum_{i=1}^{n}\frac{1}{I_{\nu_1,\tau_1}\left[\frac{p_1\nu_1-1}{p_1\nu_1+p_1\tau_1}\right]}\frac{\partial I_{\nu_1,\tau_1}\left[\frac{p_1\nu_1-1}{p_1\nu_1+p_1\tau_1}\right]}{\partial \nu_1}\nonumber\\
&+I\left[y_i > u(\bm{x}_i;\bm{\beta})\right]\sum_{i=1}^{n}\frac{1}{r(\bm{\alpha})-1}\frac{\partial r(\bm{\alpha})}{\partial \nu_1}\nonumber,
\end{align}
\begin{align}
\frac{\partial \ell(\bm{\beta}, \bm{\alpha};\bm{y})}{\partial \nu_2}
&=I\left[y_i\le u(\bm{x}_i;\bm{\beta})\right]\sum_{i=1}^{n}\frac{1}{r(\bm{\alpha})}\frac{\partial r(\bm{\alpha})}{\partial \nu_2}\nonumber\\
&+I\left[y_i > u(\bm{x}_i;\bm{\beta})\right]\sum_{i=1}^{n}\frac{1}{r(\bm{\alpha})-1}\frac{\partial r(\bm{\alpha})}{\partial \nu_2}\nonumber\\
&+I\left[y_i > u(\bm{x}_i;\bm{\beta})\right]\sum_{i=1}^{n}\frac{1}{f_{\text{GBII}}(y_i;\mu_{2}(\bm{x}_i;\bm{\beta}), p_2,\nu_2,\tau_2)}\frac{\partial f_{\text{GBII}}(y_i;\mu_{2}(\bm{x}_i;\bm{\beta}), p_2,\nu_2,\tau_2)}{\partial \nu_2}\nonumber\\
&+I\left[y_i > u(\bm{x}_i;\bm{\beta})\right]\sum_{i=1}^{n}\frac{1}{1-I_{\nu_2,\tau_2}\left[\frac{p_2\nu_2-1}{p_2\nu_2+p_2\tau_2}\right]}\frac{\partial I_{\nu_2,\tau_2}\left[\frac{p_2\nu_2-1}{p_2\nu_2+p_2\tau_2}\right]}{\partial \nu_2}\nonumber.
\end{align}


Note that the gradients of the $r(\bm{\alpha})$, $f_{\text{GBII}}(y_i;\mu_j(\bm{x}_i;\bm{\beta}), p_j,\nu_j,\tau_j)$ and $I_{\nu_j,\tau_j}\left[\frac{p_j\nu_j-1}{p_j\nu_j+p_j\tau_j}\right]$ for $j=1,2$ 
can be calculated based on the following equations.
For simplicity,
the subscripts of $I_{\nu_j,\tau_j}(\pi_j)$ and $f_{\text{GBII}}(y_i;\mu_j(\bm{x}_i;\bm{\beta}), p_j,\nu_j,\tau_j)$ are omitted. 

First, the partial derivatives of $r(\bm{\alpha})$ are given by:
\begin{align}
\frac{\partial r(\bm{\alpha})}{\partial p_1}
&=\frac{\frac{\partial I_{\nu_1,\tau_1}(\pi_1)}{\partial p_1}\omega-I_{\nu_1,\tau_1}(\pi_1)
	\left[\frac{\partial I_{\nu_1,\tau_1}(\pi_1)}{\partial p_1}
	+\frac{\partial \phi}{\partial p_1}(1-I_{\nu_2,\tau_2}(\pi_2))\right]}{\omega^{2}}\nonumber,\\
\frac{\partial r(\bm{\alpha})}{\partial p_2}
&=\frac{-I_{\nu_1,\tau_1}(\pi_1)\left[\frac{\partial \phi}{\partial p_2}\left[1-I_{\nu_2,\tau_2}(\pi_2)\right]-\phi\frac{\partial I_{\nu_2,\tau_2}(\pi_2)}{\partial p_2}\right]}{\omega^{2}}\nonumber,\\
\frac{\partial r(\bm{\alpha})}{\partial \tau_1}
&=\frac{\frac{\partial I_{\nu_1,\tau_1}(\pi_1)}{\partial \tau_1}\omega-I_{\nu_1,\tau_1}(\pi_1)
	\left[\frac{\partial I_{\nu_1,\tau_1}(\pi_1)}{\partial \tau_1}
	+\frac{\partial \phi}{\partial \tau_1}(1-I_{\nu_2,\tau_2}(\pi_2))\right]}{\omega^{2}}\nonumber,\\
\frac{\partial r(\bm{\alpha})}{\partial \tau_2}
&=\frac{-I_{\nu_1,\tau_1}(\pi_1)\left[\frac{\partial \phi}{\partial \tau_2}\left[1-I_{\nu_2,\tau_2}(\pi_2)\right]-\phi\frac{\partial I_{\nu_2,\tau_2}(\pi_2)}{\partial \tau_2}\right]}{\omega^{2}}\nonumber,\\
\frac{\partial r(\bm{\alpha})}{\partial \nu_1}
&=\frac{\frac{\partial I_{\nu_1,\tau_1}(\pi_1)}{\partial \nu_1}\omega-I_{\nu_1,\tau_1}(\pi_1)
	\left[\frac{\partial I_{\nu_1,\tau_1}(\pi_1)}{\partial \nu_1}
	+\frac{\partial \phi}{\partial \nu_1}(1-I_{\nu_2,\tau_2}(\pi_2))\right]}{\omega^{2}}\nonumber,\\
\frac{\partial r(\bm{\alpha})}{\partial \nu_2}
&=\frac{-I_{\nu_1,\tau_1}(\pi_1)\left[\frac{\partial \phi}{\partial \nu_2}\left[1-I_{\nu_2,\tau_2}(\pi_2)\right]-\phi\frac{\partial I_{\nu_2,\tau_2}(\pi_2)}{\partial \nu_2}\right]}{\omega^{2}}\nonumber,
\end{align}

where 
\begin{align}
\omega
&=I_{\nu_1,\tau_1}\left[\frac{p_1\nu_1-1}{p_1\nu_1+p_1\tau_1}\right]+\phi \left\{1 - I_{\nu_2,\tau_2}\left[\frac{p_2\nu_2-1}{p_2\nu_2+p_2\tau_2}\right]\right\}\nonumber,\\
\pi_1
&=\frac{p_1\nu_1-1}{p_1\nu_1+p_1\tau_1}\nonumber,\\
\pi_2
&=\frac{p_2\nu_2-1}{p_2\nu_2+p_2\tau_2}\nonumber,\\
\frac{\partial \phi}{\partial p_1}
&=\phi\left[\frac{1}{p_1}+\frac{\nu_1+\tau_1}{p_1(p_1\nu_1-1)(p_1\tau_1+1)}\right]\nonumber,\\
\frac{\partial \phi}{\partial p_2}
&=-\phi\left[\frac{1}{p_2}+\frac{\nu_2+\tau_2}{p_2(p_2\nu_2-1)(p_2\tau_2+1)}\right]\nonumber,\\
\frac{\partial \phi}{\partial \tau_1}
&=\phi\left[\ln\frac{p_1\tau_1+1}{p_1\nu_1+p_1\tau_1}-\frac{1}{p_1\tau_1+1}-\psi(\tau_1)+\psi(\nu_1+\tau_1)\right]\nonumber,\\
\frac{\partial \phi}{\partial \tau_2}
&=\phi\left[\ln\frac{p_2\nu_2+p_2\tau_2}{p_2\tau_2+1}+\frac{1}{p_2\tau_2+1}+\psi(\tau_2)-\psi(\nu_2+\tau_2)\right]\nonumber,\\
\frac{\partial \phi}{\partial \nu_1}
&=\phi\left[\ln\frac{p_1\nu_1-1}{p_1\nu_1+p_1\tau_1}+\frac{1}{p_1\nu_1-1}-\psi(\nu_1)+\psi(\nu_1+\tau_1)\right]\nonumber,\\
\frac{\partial \phi}{\partial \nu_2}
&=\phi\left[\ln\frac{p_2\nu_2+p_2\tau_2}{p_2\nu_2-1}-\frac{1}{p_2\nu_2-1}+\psi(\nu_2)-\psi(\nu_2+\tau_2)\right]\nonumber,
\end{align}
with $\psi(\cdot)$ denotes the Digamma function.

Then, we need to calculate the partial derivatives of $I_{\nu,\tau}(\pi)=I_{\nu,\tau}\left[\frac{p\nu-1}{p\nu+p\tau}\right]$ which are given by
\begin{align}
\frac{\partial I_{\nu,\tau}(\pi)}{\partial p}
&=\frac{\partial I_{\nu,\tau}\left[\frac{p\nu-1}{p\nu+p\tau}\right]}{\partial p}
=\frac{(p\tau+1)^{(\tau-1)}(p\nu-1)^{(\nu-1)}}{B(\nu,\tau)p(p\nu+p\tau)^{(\nu+\tau-1)}}\nonumber,\\
\frac{\partial I_{\nu,\tau}(\pi)}{\partial \tau}
&=\frac{\partial I_{\nu,\tau}\left[\frac{p\nu-1}{p\nu+p\tau}\right]}{\partial \tau}
=\frac{\Gamma(\tau)\Gamma(\nu+\tau)}{\Gamma(\nu)}\left[(1-\pi)^{\tau}{}_3\tilde{F}_2(\tau,\tau,1-\nu;\tau+1,\tau+1;1-\pi)\right]\nonumber\\
&\quad \quad \quad \quad \quad \quad \quad \quad +I_{\tau,\nu}(1-\pi)\left[\psi(\tau)-\psi(\nu+\tau)-\log(1-\pi)\right]\nonumber,\\
\frac{\partial I_{\nu,\tau}(\pi)}{\partial \nu}
&=\frac{\partial I_{\nu,\tau}\left[\frac{p\nu-1}{p\nu+p\tau}\right]}{\partial \nu}
=-\frac{\Gamma(\nu)\Gamma(\nu+\tau)}{\Gamma(\tau)}\left[\pi^{\nu}{}_3\tilde{F}_2(\nu,\nu,1-\tau;\nu+1,\nu+1;\pi)\right]\nonumber\\
&\quad \quad \quad \quad \quad \quad \quad \quad  +I_{\nu,\tau}(\pi)\left[\psi(\nu+\tau)-\psi(\nu)+\log(\pi)\right]\nonumber,
\end{align} 
where ${}_3\tilde{F}_2(\cdot)$ denotes the Regularized generalized hypergeometric function, with $\emph{alist}$ contain three-parameters and $\emph{blist}$ contain two parameters.

Finally, 
the partial derivatives of $f_{\text{GBII}}(y_i;\mu(\bm{x}_i;\bm{\beta}), p,\nu,\tau)$ with respected to $p$, $\tau$ and $\nu$ are given by
\begin{align}
\frac{\partial f_{\text{GBII}}(y_i;\mu(\bm{x}_i;\bm{\beta}), p,\nu,\tau)}{\partial p}
&=f_{\text{GBII}}(y_i;\mu(\bm{x}_i;\bm{\beta}), p,\nu,\tau)\left[\frac{1}{p}+\frac{\tau y_i^{p}\ln\frac{\mu(\bm{x}_i;\bm{\beta})}{y_i}+\nu \mu(\bm{x}_i;\bm{\beta})^{p}\ln\frac{y_i}{\mu(\bm{x}_i;\bm{\beta})}}{y_i^{p}+\mu(\bm{x}_i;\bm{\beta})^{p}}\right]\nonumber,\\
\frac{\partial f_{\text{GBII}}(y_i;\mu(\bm{x}_i;\bm{\beta}), p,\nu,\tau)}{\partial \tau}
&=f_{\text{GBII}}(y_i;\mu(\bm{x}_i;\bm{\beta}), p,\nu,\tau)\left[\ln\frac{\mu(\bm{x}_i;\bm{\beta})^{p}}{y_i^{p}+\mu(\bm{x}_i;\bm{\beta})^{p}}\right]\nonumber,\\
\frac{\partial f_{\text{GBII}}(y_i;\mu(\bm{x}_i;\bm{\beta}), p,\nu,\tau)}{\partial \nu}
&=f_{\text{GBII}}(y_i;\mu(\bm{x}_i;\bm{\beta}), p,\nu,\tau)\left[\ln\frac{y_i^{p}}{y_i^{p}+\mu(\bm{x}_i;\bm{\beta})^{p}}\right]\nonumber.
\end{align} 

{{
Note that $\sigma_1=p_1\tau_1, \sigma_2=p_2\tau_2, \delta_1 = p_1\nu_1, \delta_2=p_2\nu_2$, we can obtain  the first order partial derivatives of the log-likelihood function $ \ell(\bm{\beta}, \bm{\alpha};\bm{y}) $, with respect to  $\bm{\alpha}$, which are give by:
\begin{align*}
\frac{\partial \ell(\bm{\beta}, \bm{\alpha};\bm{y})}{\partial \alpha_1}&=p_1\frac{\partial \ell(\bm{\beta}, \bm{\alpha};\bm{y})}{\partial p_1},\\
\frac{\partial \ell(\bm{\beta}, \bm{\alpha};\bm{y})}{\partial \alpha_2}&=p_2\frac{\partial \ell(\bm{\beta}, \bm{\alpha};\bm{y})}{\partial p_2},\\
\frac{\partial \ell(\bm{\beta}, \bm{\alpha};\bm{y})}{\partial \alpha_3}&=\frac{1}{2}\left[\tau_1\frac{\partial \ell(\bm{\beta}, \bm{\alpha};\bm{y})}{\partial \tau_1}+p_1\frac{\partial \ell(\bm{\beta}, \bm{\alpha};\bm{y})}{\partial p_1}\right],\\
\frac{\partial \ell(\bm{\beta}, \bm{\alpha};\bm{y})}{\partial \alpha_4}&=\frac{1}{2}\left[\tau_2\frac{\partial \ell(\bm{\beta}, \bm{\alpha};\bm{y})}{\partial \tau_2}+p_2\frac{\partial \ell(\bm{\beta}, \bm{\alpha};\bm{y})}{\partial p_2}\right],\\
\frac{\partial \ell(\bm{\beta}, \bm{\alpha};\bm{y})}{\partial \alpha_5}&=\frac{1}{2}\left[\nu_1\frac{\partial \ell(\bm{\beta}, \bm{\alpha};\bm{y})}{\partial \nu_1}+p_1\frac{\partial \ell(\bm{\beta}, \bm{\alpha};\bm{y})}{\partial p_1}\right],\\
\frac{\partial \ell(\bm{\beta}, \bm{\alpha};\bm{y})}{\partial \alpha_6}&=\frac{1}{2}\left[\nu_2\frac{\partial \ell(\bm{\beta}, \bm{\alpha};\bm{y})}{\partial \nu_2}+p_2\frac{\partial \ell(\bm{\beta}, \bm{\alpha};\bm{y})}{\partial p_2}\right].
\end{align*}}}

The first order partial derivatives of the log-likelihood function $\partial \ell(\bm{\beta}, \bm{\alpha};\bm{y}) $, with respect to  $\bm{\beta}$ are give by:
\begin{align}
\frac{\partial \ell(\bm{\beta}, \bm{\alpha};\bm{y})}{\partial \bm{\beta}}
&=I\left[y_i\le u(\bm{x}_i;\bm{\beta})\right]\sum_{i=1}^{n}\frac{1}{f_{\text{GBII}}(y_i;\mu_{1}(\bm{x}_i;\bm{\beta}), p_1,\nu_1,\tau_1)}\frac{\partial f_{\text{GBII}}(y_i;\mu_{1}(\bm{x}_i;\bm{\beta}), p_1,\nu_1,\tau_1)}{\partial \mu_{1}(\bm{x}_i;\bm{\beta})}\frac{\partial \mu_{1}(\bm{x}_i;\bm{\beta})}{\partial \bm{\beta}}\nonumber\\
&+I\left[y_i > u(\bm{x}_i;\bm{\beta})\right]\sum_{i=1}^{n}\frac{1}{f_{\text{GBII}}(y_i;\mu_{2}(\bm{x}_i;\bm{\beta}), p_2,\nu_2,\tau_2)}\frac{\partial f_{\text{GBII}}(y_i;\mu_{2}(\bm{x}_i;\bm{\beta}), p_2,\nu_2,\tau_2)}{\partial \mu_{2}(\bm{x}_i;\bm{\beta})}\frac{\partial \mu_{2}(\bm{x}_i;\bm{\beta})}{\partial \bm{\beta}}\nonumber,
\end{align}
where
\begin{align}
\frac{\partial f_{\text{GBII}}(y_i;\mu(\bm{x}_i;\bm{\beta}), p,\nu,\tau)}{\partial \bm{\beta}}
&=\frac{\partial f_{\text{GBII}}(y_i;\mu(\bm{x}_i;\bm{\beta}), p,\nu,\tau)}{\partial \mu(\bm{x}_i;\bm{\beta})}\frac{\partial \mu(\bm{x}_i;\bm{\beta})}{\partial \bm{\beta}}\nonumber\\
&=f_{\text{GBII}}(y_i;\mu(\bm{x}_i;\bm{\beta}), p,\nu,\tau)\frac{p\left[\tau y_i^{p}-\nu \mu(\bm{x}_i;\bm{\beta})^{p}\right]}{\mu(\bm{x}_i;\bm{\beta})\left[y_i^{p}+\mu(\bm{x}_i;\bm{\beta})^{p}\right]}\frac{\partial \mu(\bm{x}_i;\bm{\beta})}{\partial \bm{\beta}}\nonumber.
\end{align}

\clearpage

\subsection{Distribution nested within the GBII distribution}\label{app:sub-GBII}

\begin{table}[hb]
	\centering
	\caption{Distribution nested within the GBII distribution}
	\begin{tabular*}{\hsize}{@{}@{\extracolsep{\fill}}lllll@{}}
		\toprule
		Distribution&Npars &\multicolumn{3}{c}{Density function}\\
		\hline
		GBII & 4&  & $\text{GBII}(p,\mu,\nu,\tau)$&$\frac{\abs{p}}{B(\nu,\tau)y}\frac{\mu^{p \tau}y^{p\nu}}{\left(y^p+\mu^p\right)^{\nu+\tau}}$\\
		\tabincell{l}{Beta distribution \\ of the second kind} & 3& $\text{BII}(\mu,\nu,\tau)$ & $\text{GBII}(1,\mu,\nu,\tau)$&$\frac{1}{B(\nu,\tau)y}\frac{\mu^{ \tau}y^{\nu}}{\left(y+\mu\right)^{\nu+\tau}}$\\
		Burr& 3& $\text{B}(p,\mu,\tau)$ & $\text{GBII}(p,\mu,1,\tau)$&
		$\frac{p\tau\mu^{p\tau}y^{p-1}}{\left(\mu^p+y^p\right)^{\tau+1}}$ \\
		Inverse Burr&3&$\text{IB}(p,\nu,\tau)$&$\text{GBII}(p,\mu,\nu,1)$&
		$\frac{p\nu\mu^{p}y^{\nu p-1}}{\left(\mu^p+y^p\right)^{\nu+1}}$\\
		GLMGA&3&$\text{G}(p,\mu,\nu)$&$\text{GBII}(p,\mu,\nu,\frac{1}{2})$&$\frac{{p}}{B(\nu,\frac{1}{2})y}\frac{\mu^{ \frac{p}{2}}y^{p\nu}}{\left(y^p+\mu^p\right)^{\nu+\frac{1}{2}}}$\\
				\tabincell{l}{Inverse\\GLMGA}&3&$\text{IG}(p,\mu,\tau)$&$\text{GBII}(p,\mu,\frac{1}{2},\tau)$&$\frac{{p}}{B(\frac{1}{2},\tau)y}\frac{\mu^{ {p}\tau}y^{p\frac{1}{2}}}{\left(y^p+\mu^p\right)^{\frac{1}{2}+\tau}}$\\
		Paralogistic & 2 & $\text{P}(p,\mu)$ &$\text{GBII}(p,\mu, 1,p)$ & $\frac{p^2\mu^{p^2}y^{p-1}}{\left(\mu^p+y^{p}\right)^{p+1}}$\\
		\tabincell{l}{Inverse\\Paralogistic}& 2&$\text{IP}(p,\mu)$ &$\text{GBII}(p,\mu, p, 1)$&
		$\frac{p^2\mu^{p}y^{p^2-1}}{\left(\mu^p+y^p\right)^{p+1}}$
		\\
		\bottomrule
	\end{tabular*}%
\end{table}%

{
\subsection{Re-parameterized composite GBII distributions}\label{app:re-parameterized-GBII}
\begin{table}[hb]
	\centering
	\caption{The re-parameterized composite GBII distributions used for estimation.}
	\renewcommand\arraystretch{1.5} 
	\begin{tabular*}{\hsize}{@{}@{\extracolsep{\fill}}ccc@{}}
		\toprule
		Npars & parameterized distribution with $p, \mu, \tau, \nu$ &	re-parameterized distribution with $p, \mu, \delta, \sigma$ \\
		\hline
	7	& ComGBII $(\mu_2, p_1, p_2, \tau_1, \tau_2, \nu_1, \nu_2)$& ComGBII $(\mu_2, p_1, p_2, \sigma_1, \sigma_2, \delta_1, \delta_2)$ \\
	6	 & GBIIG $( \mu_2, p_1, p_2, \tau_1, \tau_2, \nu_1, \nu_2 = \frac{1}{2})$&GBIIG $(\mu_2, p_1, p_2, \sigma_1, \sigma_2, \delta_1, \delta_2={p_2}/{2})$\\
	5	 & BIIG  $( \mu_2, p_1=1, p_2, \tau_1, \tau_2, \nu_1, \nu_2 = \frac{1}{2})$& BIIG $(\mu_2, p_1=1, p_2, \sigma_1, \sigma_2, \delta_1, \delta_2={p_2}/{2})$\\
	5	&  BG  $( \mu_2, p_1, p_2, \tau_1, \tau_2, \nu_1=1, \nu_2 = \frac{1}{2})$& BG $(\mu_2, p_1, p_2, \sigma_1, \sigma_2, \delta_1=p_1, \delta_2={p_2}/{2})$\\
	5	&  IBG  $( \mu_2, p_1, p_2, \tau_1=1, \tau_2, \nu_1, \nu_2 = \frac{1}{2})$& IBG $(\mu_2, p_1, p_2, \sigma_1=p_1, \sigma_2, \delta_1, \delta_2={p_2}/{2})$\\
	4	&  PG  $(\mu_2, p_1, p_2, \tau_1=p_1, \tau_2, \nu_1=1, \nu_2=\frac{1}{2})$& PG $(\mu_2, p_1, p_2, \sigma_1=p_1^2, \sigma_2, \delta_1=p_1, \delta_2={p_2}/{2})$\\
	4	&  IPG  $(\mu_2, p_1, p_2, \tau_1=1, \tau_2,   \nu_1=p_1 ,  \nu_2=\frac{1}{2})$& IPG $(\mu_2, p_1, p_2, \sigma_1=p_1, \sigma_2, \delta_1=p_1^2, \delta_2={p_2}/{2})$\\
		\bottomrule
	\end{tabular*}%
		\begin{tablenotes}
	\item \small Note that $\sigma_1=p_1\tau_1, \sigma_2=p_2\tau_2, \delta_1 = p_1\nu_1, \delta_2=p_2\nu_2$.
\end{tablenotes}
\end{table}%
}

\clearpage
\clearpage
\bibliographystyle{plainnat}
\bibliography{mybibfile}

\begin{thebibliography}{43}
\providecommand{\natexlab}[1]{#1}
\providecommand{\url}[1]{\texttt{#1}}
\expandafter\ifx\csname urlstyle\endcsname\relax
  \providecommand{\doi}[1]{doi: #1}\else
  \providecommand{\doi}{doi: \begingroup \urlstyle{rm}\Url}\fi

\bibitem[Azzalini et~al.(2002)Azzalini, Del~Cappello, Kotz,
  et~al.]{azzalini2002log}
Adelchi Azzalini, Thomas Del~Cappello, Samuel Kotz, et~al.
\newblock Log-skew-normal and log-skew-t distributions as models for family
  income data.
\newblock \emph{Journal of Income Distribution}, 11\penalty0 (3):\penalty0
  12--20, 2002.

\bibitem[Baione and Biancalana(2019)]{baione2019individual}
Fabio Baione and Davide Biancalana.
\newblock An individual risk model for premium calculation based on quantile: a
  comparison between generalized linear models and quantile regression.
\newblock \emph{North American Actuarial Journal}, 23\penalty0 (4):\penalty0
  573--590, 2019.

\bibitem[Bakar et~al.(2015)Bakar, Hamzah, Maghsoudi, and
  Nadarajah]{bakar2015modeling}
SA~Abu Bakar, NA~Hamzah, M~Maghsoudi, and S~Nadarajah.
\newblock Modeling loss data using composite models.
\newblock \emph{Insurance: Mathematics and Economics}, 61:\penalty0 146--154,
  2015.

\bibitem[Beirlant et~al.(1998)Beirlant, Goegebeur, Verlaak, and
  Vynckier]{beirlant1998burr}
Jan Beirlant, Yuri Goegebeur, Robert Verlaak, and Petra Vynckier.
\newblock Burr regression and portfolio segmentation.
\newblock \emph{Insurance: Mathematics and Economics}, 23\penalty0
  (3):\penalty0 231--250, 1998.

\bibitem[Beirlant et~al.(2004)Beirlant, Goegebeur, Segers, and
  Teugels]{beirlant2004}
Jan Beirlant, Yuri Goegebeur, Johan Segers, and Jozef Teugels.
\newblock \emph{Statistics of Extremes: Theory and Applications}.
\newblock Wiley Series in Probability and Statistics, 2004.

\bibitem[Bernardi et~al.(2012)Bernardi, Maruotti, and
  Petrella]{bernardi2012skew}
Mauro Bernardi, Antonello Maruotti, and Lea Petrella.
\newblock Skew mixture models for loss distributions: a bayesian approach.
\newblock \emph{Insurance: Mathematics and Economics}, 51\penalty0
  (3):\penalty0 617--623, 2012.

\bibitem[Bhati and Ravi(2018)]{bhati2018generalized}
Deepesh Bhati and Sreenivasan Ravi.
\newblock On generalized log-moyal distribution: A new heavy tailed size
  distribution.
\newblock \emph{Insurance: Mathematics and Economics}, 79:\penalty0 247--259,
  2018.

\bibitem[Bladt(2022)]{bladt2021phase}
Martin Bladt.
\newblock Phase-type distributions for claim severity regression modeling.
\newblock \emph{ASTIN Bulletin: The Journal of the IAA}, pages 1--32, 2022.

\bibitem[Calderín-Ojeda and Kwok(2016)]{calderin2016modeling}
Enrique Calderín-Ojeda and Chun~Fung Kwok.
\newblock Modeling claims data with composite stoppa models.
\newblock \emph{Scandinavian Actuarial Journal}, 2016\penalty0 (9):\penalty0
  817--836, 2016.

\bibitem[Chan et~al.(2018)Chan, Choy, Makov, and Landsman]{chan2018modelling}
JSK Chan, STB Choy, UE~Makov, and Z~Landsman.
\newblock Modelling insurance losses using contaminated generalised beta
  type-{II} distribution.
\newblock \emph{ASTIN Bulletin: The Journal of the IAA}, 48\penalty0
  (2):\penalty0 871--904, 2018.

\bibitem[Cooray and Ananda(2005)]{cooray2005modeling}
Kahadawala Cooray and Malwane~MA Ananda.
\newblock Modeling actuarial data with a composite lognormal-pareto model.
\newblock \emph{Scandinavian Actuarial Journal}, 2005\penalty0 (5):\penalty0
  321--334, 2005.

\bibitem[del Castillo et~al.(2017)del Castillo, Daoudi, and Serra]{del2017full}
Joan del Castillo, Jalila Daoudi, and Isabel Serra.
\newblock The full tails gamma distribution applied to model extreme values.
\newblock \emph{ASTIN Bulletin: The Journal of the IAA}, 47\penalty0
  (3):\penalty0 895--917, 2017.

\bibitem[Dong and Chan(2013)]{dong2013bayesian}
Alice~XD Dong and JSK Chan.
\newblock Bayesian analysis of loss reserving using dynamic models with
  generalized beta distribution.
\newblock \emph{Insurance: Mathematics and Economics}, 53\penalty0
  (2):\penalty0 355--365, 2013.

\bibitem[Dunn and Smyth(1996)]{dunn1996randomized}
Peter~K Dunn and Gordon~K Smyth.
\newblock Randomized quantile residuals.
\newblock \emph{Journal of Computational and Graphical Statistics}, 5\penalty0
  (3):\penalty0 236--244, 1996.

\bibitem[Fissler et~al.(2023)Fissler, Merz, and W{\"u}thrich]{fissler2021deep}
Tobias Fissler, Michael Merz, and Mario~V W{\"u}thrich.
\newblock Deep quantile and deep composite triplet regression.
\newblock \emph{Insurance: Mathematics and Economics}, 109:\penalty0 94--112,
  2023.

\bibitem[Frees et~al.(2011)Frees, Meyers, and Cummings]{frees2011summarizing}
Edward~W Frees, Glenn Meyers, and A~David Cummings.
\newblock Summarizing insurance scores using a gini index.
\newblock \emph{Journal of the American Statistical Association}, 106\penalty0
  (495):\penalty0 1085--1098, 2011.

\bibitem[Frees et~al.(2014)Frees, Meyers, and Cummings]{frees2014insurance}
Edward~W Frees, Glenn Meyers, and A~David Cummings.
\newblock Insurance ratemaking and a gini index.
\newblock \emph{Journal of Risk and Insurance}, 81\penalty0 (2):\penalty0
  335--366, 2014.

\bibitem[Fung et~al.(2019)Fung, Badescu, and Lin]{fung2019class}
Tsz~Chai Fung, Andrei~L Badescu, and X~Sheldon Lin.
\newblock A class of mixture of experts models for general insurance:
  Theoretical developments.
\newblock \emph{Insurance: Mathematics and Economics}, 89:\penalty0 111--127,
  2019.

\bibitem[Fung et~al.(2022)Fung, Badescu, and Lin]{fung2022fitting}
Tsz~Chai Fung, Andrei~L Badescu, and X~Sheldon Lin.
\newblock Fitting censored and truncated regression data using the mixture of
  experts models.
\newblock \emph{North American Actuarial Journal}, 26\penalty0 (4):\penalty0
  496--520, 2022.

\bibitem[Fung et~al.(2023)Fung, Tzougas, and W{\"u}thrich]{fung2021mixture}
Tsz~Chai Fung, George Tzougas, and Mario~V W{\"u}thrich.
\newblock Mixture composite regression models with multi-type feature
  selection.
\newblock \emph{North American Actuarial Journal}, 27\penalty0 (2):\penalty0
  396--428, 2023.

\bibitem[Gan and Valdez(2018)]{gan2018fat}
Guojun Gan and Emiliano~A Valdez.
\newblock Fat-tailed regression modeling with spliced distributions.
\newblock \emph{North American Actuarial Journal}, 22\penalty0 (4):\penalty0
  554--573, 2018.

\bibitem[Gill et~al.(2005)Gill, Murray, and Saunders]{gill2005snopt}
Philip~E Gill, Walter Murray, and Michael~A Saunders.
\newblock Snopt: An sqp algorithm for large-scale constrained optimization.
\newblock \emph{SIAM review}, 47\penalty0 (1):\penalty0 99--131, 2005.

\bibitem[G{\'o}mez-D{\'e}niz et~al.(2013)G{\'o}mez-D{\'e}niz,
  Calder{\'\i}n-Ojeda, and Sarabia]{gomez2013gamma}
Emilio G{\'o}mez-D{\'e}niz, Enrique Calder{\'\i}n-Ojeda, and
  Jos{\'e}~Mar{\'\i}a Sarabia.
\newblock Gamma-generalized inverse gaussian class of distributions with
  applications.
\newblock \emph{Communications in Statistics-Theory and Methods}, 42\penalty0
  (6):\penalty0 919--933, 2013.

\bibitem[Grün and Miljkovic(2019)]{grun2019extending}
Bettina Grün and Tatjana Miljkovic.
\newblock Extending composite loss models using a general framework of advanced
  computational tools.
\newblock \emph{Scandinavian Actuarial Journal}, 2019\penalty0 (8):\penalty0
  642--660, 2019.

\bibitem[Klein et~al.(2014)Klein, Denuit, Lang, and Kneib]{klein2014nonlife}
Nadja Klein, Michel Denuit, Stefan Lang, and Thomas Kneib.
\newblock Nonlife ratemaking and risk management with bayesian generalized
  additive models for location, scale, and shape.
\newblock \emph{Insurance: Mathematics and Economics}, 55:\penalty0 225--249,
  2014.

\bibitem[Klugman et~al.(2012)Klugman, Panjer, and Willmot]{klugman2012loss}
Stuart~A Klugman, Harry~H Panjer, and Gordon~E Willmot.
\newblock \emph{Loss models: from data to decisions}, volume 715.
\newblock John Wiley \& Sons, 2012.

\bibitem[Landsman et~al.(2016)Landsman, Makov, and Shushi]{landsman2016tail}
Zinoviy Landsman, Udi Makov, and Tomer Shushi.
\newblock Tail conditional moments for elliptical and log-elliptical
  distributions.
\newblock \emph{Insurance: Mathematics and Economics}, 71:\penalty0 179--188,
  2016.

\bibitem[Laudag{\'e} et~al.(2019)Laudag{\'e}, Desmettre, and
  Wenzel]{laudage2019severity}
Christian Laudag{\'e}, Sascha Desmettre, and J{\"o}rg Wenzel.
\newblock Severity modeling of extreme insurance claims for tariffication.
\newblock \emph{Insurance: Mathematics and Economics}, 88:\penalty0 77--92,
  2019.

\bibitem[Li et~al.(2021)Li, Beirlant, and Meng]{li2021generalizing}
Zhengxiao Li, Jan Beirlant, and Shengwang Meng.
\newblock Generalizing the log-moyal distribution and regression models for
  heavy-tailed loss data.
\newblock \emph{ASTIN Bulletin: The Journal of the IAA}, 51\penalty0
  (1):\penalty0 57--99, 2021.

\bibitem[McDonald and Xu(1995)]{mcdonald1995generalization}
James~B McDonald and Yexiao~J Xu.
\newblock A generalization of the beta distribution with applications.
\newblock \emph{Journal of Econometrics}, 66\penalty0 (1-2):\penalty0 133--152,
  1995.

\bibitem[Miljkovic and Gr{\"u}n(2016)]{miljkovic2016modeling}
Tatjana Miljkovic and Bettina Gr{\"u}n.
\newblock Modeling loss data using mixtures of distributions.
\newblock \emph{Insurance: Mathematics and Economics}, 70:\penalty0 387--396,
  2016.

\bibitem[Nadarajah and Bakar(2014)]{nadarajah2014}
S.~Nadarajah and S.~A.~A. Bakar.
\newblock New composite models for the danish fire insurance data.
\newblock \emph{Scandinavian Actuarial Journal}, 2014\penalty0 (2):\penalty0
  180--187, 2014.

\bibitem[Nocedal and Wright(2006)]{nocedal2006numerical}
Jorge Nocedal and Stephen Wright.
\newblock \emph{Numerical optimization}.
\newblock Springer Science \& Business Media, 2006.

\bibitem[Oehlert(1992)]{oehlert1992note}
Gary~W Oehlert.
\newblock A note on the delta method.
\newblock \emph{The American Statistician}, 46\penalty0 (1):\penalty0 27--29,
  1992.

\bibitem[Punzo et~al.(2018)Punzo, Bagnato, and Maruotti]{punzo2018compound}
Antonio Punzo, Luca Bagnato, and Antonello Maruotti.
\newblock Compound unimodal distributions for insurance losses.
\newblock \emph{Insurance: Mathematics and Economics}, 81:\penalty0 95--107,
  2018.

\bibitem[Reynkens et~al.(2017)Reynkens, Verbelen, Beirlant, and
  Antonio]{reynkens2017modelling}
Tom Reynkens, Roel Verbelen, Jan Beirlant, and Katrien Antonio.
\newblock Modelling censored losses using splicing: A global fit strategy with
  mixed erlang and extreme value distributions.
\newblock \emph{Insurance: Mathematics and Economics}, 77:\penalty0 65--77,
  2017.

\bibitem[Scollnik(2007)]{scollnik2007composite}
David~PM Scollnik.
\newblock On composite lognormal-pareto models.
\newblock \emph{Scandinavian Actuarial Journal}, 2007\penalty0 (1):\penalty0
  20--33, 2007.

\bibitem[Scollnik and Sun(2012)]{scollnik2012modeling}
David~PM Scollnik and Chenchen Sun.
\newblock Modeling with weibull-pareto models.
\newblock \emph{North American Actuarial Journal}, 16\penalty0 (2):\penalty0
  260--272, 2012.

\bibitem[Shi and Yang(2018)]{shi2018pair}
Peng Shi and Lu~Yang.
\newblock Pair copula constructions for insurance experience rating.
\newblock \emph{Journal of the American Statistical Association}, 113\penalty0
  (521):\penalty0 122--133, 2018.

\bibitem[Tzougas and Karlis(2020)]{tzougas2020algorithm}
George Tzougas and Dimitris Karlis.
\newblock An {EM} algorithm for fitting a new class of mixed exponential
  regression models with varying dispersion.
\newblock \emph{ASTIN Bulletin: The Journal of the IAA}, 50\penalty0
  (2):\penalty0 555--583, 2020.

\bibitem[Verbelen et~al.(2015)Verbelen, Gong, Antonio, Badescu, and
  Lin]{verbelen2015fitting}
Roel Verbelen, Lan Gong, Katrien Antonio, Andrei Badescu, and Sheldon Lin.
\newblock Fitting mixtures of erlangs to censored and truncated data using the
  {EM} algorithm.
\newblock \emph{ASTIN Bulletin: The Journal of the IAA}, 45\penalty0
  (3):\penalty0 729--758, 2015.

\bibitem[Yang et~al.(2018)Yang, Qian, and Zou]{yang2018insurance}
Yi~Yang, Wei Qian, and Hui Zou.
\newblock Insurance premium prediction via gradient tree-boosted tweedie
  compound poisson models.
\newblock \emph{Journal of Business \& Economic Statistics}, 36\penalty0
  (3):\penalty0 456--470, 2018.

\bibitem[Ye(1987)]{ye1987interior}
Yinyu Ye.
\newblock \emph{Interior algorithms for linear, quadratic, and linearly
  constrained non-linear programming}.
\newblock PhD thesis, Ph. D. thesis, Department of ESS, Stanford University,
  1987.

\end{thebibliography}
\end{document}